\documentclass[11pt,sort&compress]{elsarticle}
\usepackage[paper=letterpaper,top=24mm, bottom=26mm, left=26mm, right=26mm]{geometry}
\makeatletter
\def\ps@pprintTitle{%
 \let\@oddhead\@empty
 \let\@evenhead\@empty
 \def\@oddfoot{\centerline{\thepage}}%
 \let\@evenfoot\@oddfoot}
\makeatother
\usepackage{amsmath}
\usepackage{amsfonts}
\usepackage{accents}
\usepackage{amssymb}
\usepackage{mathrsfs}
\usepackage{mathtools}
\usepackage[LGR,T1]{fontenc}
\usepackage[latin1]{inputenc}
\let\oldbibliography\thebibliography
\renewcommand{\thebibliography}[1]{%
  \oldbibliography{#1}%
  \setlength{\itemsep}{1.4pt}%
}
\usepackage[cal=boondox,calscaled=1]{mathalfa}
\DeclareMathAlphabet{\bbvar}{U}{BOONDOX-ds}{m}{n}
\DeclareMathAlphabet{\bbgreek}{U}{bbold}{m}{n}
\usepackage{psfrag}
\usepackage{pstool}
\usepackage{caption}
\usepackage{tabularx}
\usepackage{multirow}
\usepackage{hhline}
\usepackage{ltablex}
\usepackage{pbox}
\usepackage{graphicx}%

\newcommand{\hook}{\text{\large{$\lrcorner$}}}

\usepackage[colorlinks=true,
            linkcolor=blue,
            urlcolor=blue,
            citecolor=blue]{hyperref}
\usepackage[hyperref]{xcolor}
\definecolor{darkred}{rgb}{.95,.0,.0}

\newcommand{\di}{\mathrm{d}}
\usepackage{tensor}
\newcommand{\ou}[3]{\tensor{#1}{^{#2}_{#3}}}
\newcommand{\uo}[3]{\tensor{#1}{_{#2}^{#3}}}

\newcommand{\I}{\mathrm{i}} 
\newcommand{\E}{\mathrm{e}} 
\newcommand{\ellp}{\ell_{\mathrm{P}}} 
\newcommand{\CC}{\mathrm{cc.}} 
\newcommand{\C}{\mathbb{C}}

\newcommand{\R}{\mathbb{R}}

\newenvironment{subalign}{\subequations\align}{\endalign\endsubequations}
\newcommand{\eref}[1]{(\ref{#1})}

\newcommand{\mtext}[1]{\text{\it #1}}
\usepackage{tikz-cd}

\newcommand{\Neq}{\mathrel{\widehat{=}}}

\newcommand\vpm{\mathbin{\vcenter{\hbox{
  \oalign{\hfil$\scriptstyle+$\hfil\cr
          \noalign{\kern-.3ex}
          $\scriptscriptstyle({-})$\cr}}}}}
\DeclareMathAlphabet{\sfit}{OT1}{fos}{sb}{it}
\DeclareMathAlphabet{\mathsf}{OT1}{fos}{sb}{n}

\definecolor{darkgreen}{rgb}{0.01, 0.75, 0.24}
\definecolor{darkblue}{rgb}{0.01, 0.24, 0.75}

\usepackage[multiple, flushmargin]{footmisc}

\let\originalleft\left
\let\originalright\right
\renewcommand{\left}{\mathopen{}\mathclose\bgroup\originalleft}
\renewcommand{\right}{\aftergroup\egroup\originalright}

\newcommand{\wtilde}[1]{\widetilde{#1}}
\newcommand{\dbarvar}{{\mathrm{d}\mkern-7.5mu\lower.18ex\hbox{$\textasciitilde$}\mkern-1.5mu}}
\newcommand{\dbar}{\operatorname{\dbarvar}}

\begin{document}

\begin{abstract}

\noindent  
\noindent In a region with a boundary, the gravitational phase space consists of radiative modes in the interior and edge modes at the boundary. Such edge modes are necessary to explain how the region couples to its environment. In this paper, we characterise the edge modes and radiative modes on a null surface for the tetradic Palatini--Holst action. Our starting point is the definition of the action and its boundary terms. We choose the least restrictive boundary conditions possible. The fixed boundary data consists of the radiative modes alone (two degrees of freedom per point). All other boundary fields are dynamical. We introduce the covariant phase space and explain how the Holst term alters the boundary symmetries. To infer the Poisson brackets among Dirac observables, we define an auxiliary phase space, where the $SL(2,\R)$ symmetries of the boundary fields is manifest. We identify the gauge generators and second-class constraints that remove the auxiliary variables. All gauge generators are at most quadratic in the fundamental $SL(2,\R)$ variables on phase space. We compute the Dirac bracket and identify the Dirac observables on the light cone. 
 Finally, we discuss various truncations to quantise the system in an effective way.

\end{abstract}%
\title{Gravitational $SL(2,\R)$ Algebra on the Light Cone}
\author{Wolfgang Wieland}
\address{Institute for Quantum Optics and Quantum Information (IQOQI)\\Austrian Academy of Sciences\\Boltzmanngasse 3, 1090 Vienna, Austria\\{\vspace{0.5em}\normalfont 12 April 2021}
}
\maketitle
\vspace{-1.2em}
\hypersetup{
  linkcolor=black,
  urlcolor=black,
  citecolor=black
}
{\tableofcontents}
\hypersetup{
  linkcolor=black,
  urlcolor=darkred,
  citecolor=darkred,
}
\begin{center}{\noindent\rule{\linewidth}{0.4pt}}\end{center}
\section{Introduction}
\noindent The goal of this paper is to consider the gravitational phase space in a causal domain and explain how the Barbero--Immirzi parameter \cite{Barbero1994,Immirziparam} deforms the $SL(2,\R)$ symmetry of the boundary fields. Our approach will be quasi-local \cite{Szabados:2004vb,BrownYork,Dittrich:2018xuk,Dittrich:2017hnl}. This is to say that we describe the gravitational field as an open (dissipative) system in a finite box \cite{Andrade:2015fna,Wieland:2020gno}. The box has a boundary, and we choose it to be null. In this way, it is easy to characterise the outgoing radiation, which escapes through the null boundary. At the null boundary itself,  we introduce a phase space, constraints and a set of Dirac observables.

The paper consists of two parts and a conclusion. The first part is \hyperref[sec2]{Section 2}, where we introduce the covariant radiative phase space \cite{Peierls,Ashtekar:1990gc,Lee:1990nz,Iyer:1994ys,Wald:1999wa} for the parity odd Palatini--Holst \cite{holst,surholst} action in causal domains. We identify the most minimal boundary conditions along the null boundary. Two radiative modes are kept fixed, all other boundary fields are unconstrained. We discuss the resulting bulk plus boundary field theory and identify its gauge symmetries. We introduce the pre-symplectic two-form on the null boundary and explain how the Barbero--Immirzi parameter deforms the gauge symmetries. We provide a characterization for the radiative data and the corner data (edge modes) \cite{Balachandran:1994up,Strominger:1997eq,Banados:1998ta,carlipbook,Carlip:2005zn,Afshar:2016wfy,Compere:2017knf,Wieland:2018ymr,Namburi:2019qja,Wieland_2020,Freidel:2020xyx,Freidel:2020svx,Rovelli:2020mpk,Rejzner:2020xid,Barnich:2021dta} at a finite null boundary for the Palatini--Holst action. The second part of the paper is \hyperref[sec3]{Section 3}, where we introduce the Dirac observables on the null boundary. These observables characterise both gravitational radiation as well as gravitational memory \cite{PhysRevLett.67.1486,memory_frauendiener}. To compute the Poisson brackets, we introduce an auxiliary phase space, where the $SL(2,\R)$ symmetry of the boundary fields is manifest. The physical phase space is obtained via symplectic reduction. There are both first-class and second-class constraints. The first-class constraints are the gauge generators of vertical diffeomorphisms along the null generators and $U(1)$ frame rotations of the co-frame field. Besides the gauge generators, there are a few second-class constraints. To compute the Poisson brackets among Dirac observables, we introduce the Dirac bracket.   Finally, we summarise the paper and discuss various strategies to approach the system at the quantum level.

\section{Barbero--Immirzi parameter on the null cone}

\noindent In this section, we explain how the $CP$-violating Barbero--Immirzi term, which has no effect on the field equations in the bulk, deforms the boundary field theory on a null surfaces and mixes the null dilation charges with an otherwise vanishing $U(1)$ charge. 

\subsection{Boundary conditions and quasi-local graviton}\label{sec2}\noindent 
Consider first the action in the bulk,\footnote{In tetrad variables, the action is $S=\frac{1}{16\pi G}\int_{M}\Big(\tfrac{1}{2}\epsilon_{\alpha\beta\mu\nu}e^\alpha\wedge e^\beta\wedge (F^{\mu\nu}-\tfrac{\Lambda}{6}e^\mu\wedge e^\nu)-\gamma^{-1}e_\alpha\wedge e_\beta\wedge F^{\alpha\beta}\Big)$.}
\begin{equation}
S[\ou{A}{A}{Ba},e_{AA'a}]=\left[\frac{\I}{8\pi\gamma G}(\gamma+\I)\int_{\mathcal{M}}\Big(\Sigma_{AB}\wedge F^{AB}-\frac{\Lambda}{6}\Sigma_{AB}\wedge\Sigma^{AB}\Big)\right]+\CC,
\end{equation}
where $\ou{F}{A}{B}=\di \ou{A}{A}{B}+\ou{A}{A}{C}\wedge \ou{A}{C}{B}$ is the $\mathfrak{sl}(2,\C)$-valued curvature of the self-dual (complex-valued Ashtekar) connection \cite{newvariables} and $\Sigma_{AB}$ is the self-dual component of the Pleba\'{n}ski two-form $e_{AA'}\wedge e_{BB'}$, namely
\begin{equation}
e_{AA'}\wedge e_{BB'}=-\bar{\epsilon}_{A'B'}\Sigma_{AB}-\epsilon_{AB}\bar{\Sigma}_{A'B'},
\end{equation}
where $e_{AA'}$ is the co-tetrad (soldering form) in spinor notation \cite{penroserindler}. The action contains two coupling constants, namely the Barbero--Immirzi parameter $\gamma$ and a cosmological constant $\Lambda$. Newton's constant $G$ plays the role of a mere conversion factor between units of length and units of mass. In the following, we will study this action in a compact and oriented spacetime region $\mathcal{M}$, whose boundary consists of two spacelike manifolds $M_0$ and $M_1$ connected by a null surface $\mathcal{N}$, i.e.\ $\partial{\mathcal{M}}=M_1^{-1}\cup\mathcal{N} \cup M_0$. The components of the boundary inherit their orientation from the bulk. \medskip

\emph{Universal structure:} To impose boundary conditions along $\mathcal{N}$, we introduce a universal ruling, which allows us to embed $\mathcal{N}$ into an abstract and oriented three-dimensional manifold $P(\mathcal{C},\pi)$, which is equipped with a surjection $\pi:P\rightarrow \mathcal{C}$ onto a two-dimensional oriented manifold $\mathcal{C}$ (the {base} manifold) such that there are homeomorphisms $\phi:\R\times \mathcal{C}\rightarrow P$ that  satisfy $(\pi\circ\phi)(u,z)=z$ for all $(u,z)\in \R\times \mathcal{C}$. The pre-image $\pi^{-1}(z)$ of any point $z\in \mathcal{C}$ on the base manifold is the null ray $\gamma_z=\pi^{-1}(z)$. A vector field $\xi^a\in TP$ is null (lightlike), if its push-forward under $\pi_\ast:TP\rightarrow T\mathcal{C}$ vanishes, i.e.
\begin{equation}
\xi^a\in VP\Leftrightarrow \pi_\ast\xi^a =0.
\end{equation}

The null surface $\mathcal{N}$ has two disconnected spacelike boundaries $\partial\mathcal{N}=\mathcal{C}_1^{-1}\cup\mathcal{C}_0$, which are cross sections of $P$, i.e.\ $\pi(\mathcal{C}_0)=\pi(\mathcal{C}_1)=\mathcal{C}$. We introduce these boundaries to cut $\mathcal{N}$ off, before any caustics may form. The induced orientation on the boundaries is determined by $\partial\mathcal{N}=\mathcal{C}_1^{-1}\cup\mathcal{C}_0$. In addition, $\partial M_0=\mathcal{C}_0$ and  $\partial M_1=\mathcal{C}_1$. \medskip 

To introduce the  boundary conditions and the corresponding boundary field theory along $\mathcal{N}$, we proceed as in \cite{Wieland:2020gno}. First of all, we note that on a null surface, there always exists a spinor-valued two-form $\eta_{Aab}$ and  a (commuting) spinor $\ell^A$ such that
\begin{equation}
\eta_{Aab}\ell_B = \frac{1}{2\I}\epsilon_{AB}\varepsilon_{ab}+\varphi^\ast_{\mathcal{N}}\Sigma_{ABab},\label{glucond}
\end{equation}
where $\varepsilon_{ab}\in\Omega^2(\mathcal{N}:\R)$ is the area two-form on the null surface and $\varphi^\ast_{\mathcal{N}}:T^\ast \mathcal{M}\rightarrow T^\ast \mathcal{N}$ denotes the pull-back. If we introduce a second and linearly independent spinor $k^A$ on $\mathcal{N}$, which be normalised such that $\varepsilon_{BA}k^B\ell^A=k_A\ell^A=1$, then there is an associate co-basis $(k_a,m_a,\bar{m}_a)$ of $T^\ast\mathcal{N}$ such that $\eta_A$ admits the decomposition
\begin{equation}
\eta_A = (\ell_A\, k-k_A m)\wedge\bar{m},\label{etadef}
\end{equation}
The complex-valued one-form $m_a$ determines the induced signature $(0++)$ metric $q_{ab}=\varphi^\ast_{\mathcal{N}}g_{ab}$ on $\mathcal{N}$. We have, in fact,
\begin{equation}
 q_{ab}+\I\varepsilon_{ab} = 2 m_a\bar{m}_b,\label{qvol}
\end{equation}
where $\varepsilon_{ab}$ is the area two-form. Besides $m_a$ and $\bar{m}_a$, the co-basis $(k_a,m_a,\bar{m}_a)$ contains also the one-form $k_a$, which selects a specific dual null vector field. We call this vector field $l^a$ and it satisfies
\begin{equation}
k_al^a=-1,\qquad \pi_\ast l^a = 0.
\end{equation}
 In the following, we restrict ourselves to configurations where $l^a$ is always future pointing. Given the co-basis $(k_a,m_a,\bar{m}_a)$, the corresponding dual basis of $T\mathcal{N}$ is $(l^a,m^a,\bar{m}^a)$, where $m^a\bar{m}_a=1$ and \begin{equation}
m^am_a=0=m_al^a, \quad m_a\bar{m}^a=1.\label{lmzero}
\end{equation}
Notice that all fields $(l^a,k_a,m_a,\bar{m}_a)$ are intrinsic to the boundary. From \eref{glucond} and \eref{etadef}, it is also possible to infer the pull-back of the tetrad to the null surface. Details can be found in \cite{Wieland:2017zkf,Wieland:2019hkz}, the result is
\begin{equation}
\varphi^\ast_{\mathcal{N}}e_{AA'}=-\I\ell_A\bar{\ell}_{A'}k+\I\ell_A\bar{k}_{A'}\bar{m}+\I k_A\bar{\ell}_{A'}{m}.\label{tetranull}
\end{equation}

The boundary action is now constructed from the boundary fields $(\eta_{Aab},\ell^A,l^a,m_a)$ and additional auxiliary variables $\varkappa_a$, $\omega_a$ and $\ou{N}{A}{ab}$. Details of the construction can be found in \cite{Wieland:2020gno}. The combined action for the bulk plus boundary field theory is given by
\begin{align}
S[\ou{A}{A}{Ba},e_{AA'a}|\eta_{Aab}&,\ell^A,\ou{N}{A}{ab},\omega_a|\varkappa_a,l^a,m_a]=\nonumber\\\nonumber
=\frac{\I}{8\pi \gamma G}(\gamma+\I)&\bigg[\int_{\mathcal{M}}\Big(\Sigma_{AB}\wedge F^{AB}-\frac{\Lambda}{6}\Sigma_{AB}\wedge\Sigma^{AB}\Big)+\\
+\int_{\mathcal{N}}\Big(&\eta_A\wedge \big(D-\tfrac{1}{2}(\varkappa+\I\omega)\big)\ell^A-\frac{\I}{2}\omega\wedge m\wedge\bar{m}+N^A\wedge (l\hook\eta_A+\ell_A\bar{m})\Big)\bigg]+\CC,\label{actndef}
\end{align}
where $D=\varphi^\ast_{\mathcal{N}}\nabla$ is the exterior $SL(2,\C)$ covariant derivative on the null surface and $l\hook\eta_A$ denotes the interior product, i.e.\ $X\hook Y\hook \eta_A =\eta_{Aab}Y^aX^b$. The complex-valued one-form $\omega_a\in\Omega^1(\mathcal{N}:\C)$ and the spinor-valued two-form $\ou{N}{A}{ab}$ are Lagrange multipliers that impose constraints such that the one-form $\eta_{Aab}$ admits the algebraic decomposition given in \eref{etadef}. If we insert these constraints back into  \eref{actndef}, the action simplifies
\begin{equation}
S = \frac{\I}{8\pi \gamma G}(\gamma+\I)\bigg[\int_{\mathcal{M}}\Big(\Sigma_{AB}\wedge F^{AB}-\frac{\Lambda}{6}\Sigma_{AB}\wedge\Sigma^{AB}\Big)+\int_{\mathcal{N}}\eta_A\wedge\big(D-\tfrac{1}{2}\varkappa\big)\ell^A\bigg]+\CC\label{actndef2}
\end{equation}
We will see below that the one-form $\varkappa_a\in \Omega^1(\mathcal{N}:\R)$ is the $U(1)$ null surface analogue of the Ashtekar--Barbero connection \cite{Ashtekar1986,Barbero1994,Immirziparam}. Although \eref{actndef2} is much simpler than \eref{actndef}, it is often more useful for us to work with the extended action \eref{actndef}.\medskip 

The boundary conditions for the extended bulk plus boundary action \eref{actndef} are such that an equivalence class of boundary fields is kept fixed along the null boundary,
\begin{equation}
\delta\mathcal{g}=0,\quad \mathcal{g}=[\varkappa_a,l^a,m_a]/_\sim.\label{gravitondef}
\end{equation}
We will identify the underlying equivalence relations $\sim$ below. Before doing so, let us also fix the boundary conditions at the corners $\{\mathcal{C}_{0},\mathcal{C}_1\}$ of the null surface and at the spacelike disks $\{M_{0},M_1\}$, where the boundary conditions are simply given by
\begin{equation}
\forall i=0,1:\;\varphi^\ast_{M_i}\delta\ou{A}{A}{Ba}=0,\quad\delta\ell^A\big|_{\mathcal{C}_i}=0.\label{bndrycond2}
\end{equation}

Let us now return to the definition of the equivalence class \eref{gravitondef}. First of all, consider the group of \emph{vertical diffeomorphisms},
\begin{equation}
\operatorname{Diff}_0(\mathcal{N})= \big\{\varphi\in\operatorname{Diff}(\mathcal{N})\big|\pi\circ\varphi\circ\pi^{-1}=\mathrm{id}_{\mathcal{C}}\big\},\label{diff0def}
\end{equation}
which are generated by null vector fields $\xi^a\in VP\Leftrightarrow\pi_\ast\xi^a=0$ that vanish at the boundary of $\mathcal{N}$, i.e.\ $\xi^a\big|_{\mathcal{C}_0}=\xi^a\big|_{\mathcal{C}_1=0}$.

Any two configurations of $[\varkappa_a,l^a,m_a]$ that only differ by such a vertical diffeomorphism are to be identified. In other words,
\begin{equation}
\forall \varphi \in \operatorname{Diff}_0(\mathcal{N}) :[\varkappa_a,\varphi_\ast l^a,m_a]\sim[\varphi^\ast\varkappa_a,l^a,\varphi^\ast m_a].\label{diffeodef1}
\end{equation}
Next, there is a trivial shift symmetry
\begin{equation}
\forall \zeta:\mathcal{N}\rightarrow\C :[\varkappa_a,l^a,m_a]\sim[\varkappa_a+\bar{\zeta}m_a+\zeta\bar{m}_a,l^a,m_a].\label{shiftdef1}
\end{equation}
Then, there are the dilations of the null normal
\begin{equation}
\forall f:\mathcal{N}\rightarrow\R :[\varkappa_a,l^a,m_a]\sim[\varkappa_a+\partial_af,\E^fl^a, m_a].\label{dilatdef1}
\end{equation}
Finally, we have the complexified conformal transformations
\begin{equation}
\forall \lambda:\mathcal{N}\rightarrow\C:[\varkappa_a,l^a,m_a]\sim\Big[\varkappa_a-\frac{1}{\gamma}\partial_a\nu,\E^{\mu}l^a, \E^{\mu+\I\nu} m_a\Big],\label{confdef1}
\end{equation}
where $\mu$ and $\nu$ denote the real and imaginary parts of the gauge parameter $\lambda$,
\begin{equation}
\operatorname{\mathfrak{Re}}(\lambda) = \mu,\qquad \operatorname{\mathfrak{Im}}(\lambda) = \nu.\label{munudef}
\end{equation}
In addition, we will also need to impose that the gauge parameters $\nu$ and $f$ for $U(1)$ rotations and dilatations satisfy the $\gamma$-twisted boundary condition
\begin{equation}
\nu\big|_{\partial\mathcal{N}}=\gamma f\big|_{\partial\mathcal{N}}\label{nubndry}
\end{equation}
at the two corners of $\mathcal{N}$.\medskip

Equation \eref{confdef1} implies that the one-form $\mathcal{A}_a=\gamma\varkappa_a$ transforms as an abelian $U(1)$ connection. This connection is the null surface analogue of the $SU(2)$ Ashtekar--Barbero connection \cite{Ashtekar1986,Barbero1994,Immirziparam}. We will see below that there is a corresponding $U(1)$ charge. In quantum theory, this charge, which measures the area of a cross section, is quantised \cite{Wieland:2017cmf}.
\renewcommand{\arraystretch}{1.3}\setcounter{table}{-1}
\begin{table}{\small
\begin{tabularx}{\textwidth}[c]{p{0.9em} p{5em} p{14.3em} p{14.6em} X}\cline{1-5}
\multicolumn{1}{l}{}&\multicolumn{2}{l}{\raisebox{0.1em}{\it kinematical histories} }&\raisebox{0.1em}{\it physical histories} &\phantom{x}\\\hline
\parbox[t]{1em}{\multirow{2}{*}{\rotatebox[origin=c]{90}{\it bulk}}}  & $e_{AA'a}$ & {\it soldering form} & $\nabla\wedge e_{AA'}=0$\\[0.1em]
& $\ou{A}{A}{Ba}$ & {\it self-dual connection} & $\big(\ou{F}{B}{A}-\frac{\Lambda}{3}\ou{\Sigma}{B}{A}\big)\wedge e_{BA'} =0$ \\[0.1em]\hline
\parbox[t]{1em}{\multirow{4}{*}{\rotatebox[origin=c]{90}{\it boundary}}} & $\ell^A$ & {\it null flag} & $D\ell^A = +\tfrac{1}{2}\big(\varkappa+\I\omega\big)\ell^A + l\hook N^A$\\
& $\eta_{Aab}$ & {\it dual (momentum) spinor} &  $D\wedge\eta_A = - \tfrac{1}{2}\big(\varkappa+\I\omega\big)\wedge \eta_A - N_A\wedge \bar{m}$\\
& $(\omega_a,\ou{N}{A}{ab})$ & {\it Lagrange multiplier fields} & $\eta_A = (\ell_A\, k-k_A m)\wedge\bar{m}$\\
& $(\varkappa_a,l^a,m_a)$ & {\it boundary data} & $(\varphi^\ast_{\mathcal{N}}e_{AA'})\wedge\bar{m} =-\I\eta_A\bar{\ell}_{A'}$
&\\\hline
\end{tabularx}\vspace{1.2em}
}
\caption{The action is a functional on the space of kinematical histories. Its saddle points define the space of physical histories $\mathcal{H}_{\mtext{phys}}\subset\mathcal{H}_{\mtext{kin}}$, which consists of all solutions of the bulk plus boundary field equations, which are given by the Einstein equations in the bulk, and additional boundary field equations.}\label{tab1} 
\end{table}
\subsection{Boundary field equations}
\noindent The equations of motion in the bulk are the familiar $\Lambda$-vacuum Einstein equations in spin-connection variables,
\begin{subalign}
\nabla\wedge\Sigma_{AB}=0&,\label{torsionless}\\
\left(\ou{F}{A}{B}-\frac{\Lambda}{3}\ou{\Sigma}{A}{B}\right)\wedge e^{BA'}=0&,\label{EEQ}
\end{subalign}
which imply that the self-dual curvature admits the algebraic decomposition,
\begin{equation}
F_{AB}=\frac{\Lambda}{3}\Sigma_{AB}+\Psi_{ABCD}\Sigma^{CD},
\end{equation}
where $\Psi_{ABCD}$ is the Weyl spinor. For our present purpose, we also need to know the evolution of the boundary fields along the null boundary for given boundary conditions \eref{gravitondef} and \eref{bndrycond2}. The coupled bulk-boundary action  is stationary with respect to variations of the boundary spinors $(\eta_{Aab},\ell^A)$ provided the following boundary field equations are satisfied, namely
\begin{align}
D\ell^A&=+\frac{1}{2}\big(\varkappa+\I\omega\big)\ell^A+l\hook N^A,\label{bndryeq1}\\
D\wedge\eta_A&=-\frac{1}{2}\big(\varkappa+\I\omega\big)\wedge \eta_A-N_A\wedge\bar{m}.\label{bndryeq2}
\end{align}
The boundary field equations propagate the boundary spinors $(\eta_{Aab},\ell^A)$ along the null boundary. The variation of the Lagrange multipliers $\omega_a$ and $\ou{N}{A}{ab}$, on the other hand, implies that $\eta_{Aab}$ satisfies the algebraic constraint \eref{etadef}. The theory in the bulk and the field theory at the boundary are connected via the gluing conditions \eref{glucond}, which are derived as an equation of motion from the coupled bulk plus boundary action \eref{actndef}. In fact, the action \eref{actndef}  is stationary for given boundary conditions (\ref{gravitondef}, \ref{bndrycond2}) with respect to variations of the (anti-)self-dual connection, provided two distinct conditions are satisfied: namely the torsionless condition \eref{torsionless} in the bulk and the additional gluing condition \eref{glucond} that couples the boundary degrees of freedom with the field theory in the bulk. 

Besides the algebraic constraints \eref{glucond} and \eref{etadef}, there is one additional constraint. At the kinematical level, the boundary one-form $\omega_a$ is complex and $\varkappa_a$ is real. If we impose that the action should be stationary under conformal variations $\delta^{\mtext{conf}}_\mu[l^a]=\mu l^a$ and $\delta^{\mtext{conf}}_\mu[m_a]=\mu m_a$ of the boundary fields, we obtain the $\gamma$-twisted reality condition 
\begin{equation}
(\gamma+\I)\omega_a-\CC =0.
\end{equation}
This condition implies that there exists a $U(1)$ connection $\Gamma$ on $\mathcal{N}$ such that
\begin{equation}
\omega_a = \frac{\gamma-\I}{\gamma}\Gamma_a,\quad\bar{\Gamma}_a=\Gamma_a,\label{Gammadef}
\end{equation}
Therefore, the real and imaginary parts of the one-form $\omega_a$ are proportional to each other via the Barbero--Immirzi parameter $\gamma$. This observation is important for us, because it allows us to infer the non-affinity $\kappa_{(l)}$ of the null generator from the boundary data $(\varkappa_a,\omega_a,l^a)$.  
 Going back to \eref{bndryeq1}, we obtain, in fact
\begin{align}
l^bD_b(\ell^A\bar{\ell}^{A'}) = l^a\big(\varkappa_a+\operatorname{\mathfrak{Re}}(\omega_a)\big)\ell^A\bar{\ell}^{A'}.
\end{align}
We then also know that the non-affinity $\kappa_{(l)}$ of $l^a$ is difined as
\begin{equation}
l^b\nabla_bl^a = \kappa_{(l)}l^a.
\end{equation}
If the torsionless equation is satisfied, the two derivatives agree, i.e.\ $l^b\nabla_bl^a=\I\uo{e}{AA'}{a}l^bD_b(\ell^A\bar{\ell}^{A'})$. Going back to \eref{Gammadef}, we thus obtain
\begin{equation}
l^a\varkappa_a = \kappa_{(l)}-\gamma^{-1}l^a\Gamma_a.\label{nonaffnty}
\end{equation}
The $U(1)$ connection $\Gamma_a$ is the null surface analogue of the  $SU(2)$ spin-connection on a spatial hypersurface. On the other hand, $\kappa_{(l)}$ is analogous to the extrinsic curvature.\footnote{Given a triad on a spatial hypersurface, the $SU(2)$ spin connection is uniquely determined by the torsionless equation. There is no analogous statement for the $U(1)$ connection $\Gamma_a$ on $\mathcal{N}$. As a one-form on $\mathcal{N}$, this connection $\Gamma_a$ is not unique. However, this non-uniqueness is mild. In fact, if the dyad $(m,\bar{m})$ on $\mathcal{N}$ is given, the pull-back of $\Gamma$ to the null generators \emph{is} unique, see \eref{phidef} below.} The one-form $\varkappa_a$ is therefore analogous of the $SU(2)$ Ashtekar--Barbero connection, which is the sum of the intrinsic spin connection and the extrinsic curvature multiplied by the Barbero--Immirzi parameter $\gamma$.

Finally, we also have to take into account the variations of the triple $(\varkappa_a,l^a,m_a)$ subject to the boundary conditions \eref{gravitondef}. Any such variation that satisfies the boundary conditions \eref{gravitondef} is a combination of an infinitesimal and vertical diffeomorphism \eref{diffeodef1}, a shift \eref{shiftdef1} of the one-form $\varkappa_a$, an infintesimal dilatation \eref{dilatdef1} and a complexified conformal transformation \eref{confdef1}. Let us consider every such variation separately and demonstrate that the action is stationary with respect to such variations provided the bulk and boundary field equations are satisfied. Hence there are no further constraints on the space of kinematical histories other than those already given in in \hyperref[tab1]{Table 1}. The set of bulk plus boundary field equations is given by  \eref{glucond}, \eref{etadef}, \eref{torsionless}, \eref{EEQ}, \eref{bndryeq1}, \eref{bndryeq2}, \eref{Gammadef}.\medskip 

Any vertical diffeomorphism $\varphi_\xi\in\mathrm{Diff}_0(\mathcal{N})$ of the triple $(\varkappa_a,l^a,m_a)$ can be extended into a diffeomorphism $\hat{\varphi}_\xi\in\mathrm{Diff}(\mathcal{M})$ of all bulk and boundary variables. Since any such vertical diffeomorphisms reduces to the identity at the boundary of $\mathcal{N}$ (recall the conditions $\xi^a\in T\mathcal{N}$, $\pi_\ast \xi^a =0$, $\xi^a|_{\partial\mathcal{N}}=0$), any such diffeomorphism $\hat{\varphi}_\xi\in\mathrm{Diff}(\mathcal{M})$ will also preserve the boundary conditions \eref{bndrycond2}. On the other hand, it is also clear that the action \eref{actndef} is invariant under any such $\hat{\varphi}_\xi\in\mathrm{Diff}(\mathcal{M})$ anyways. At a configuration where the bulk plus boundary field equations  are satisfied, the action \eref{actndef} is therefore stationary with respect to those variations $\delta^{\mtext{diffeo}}_\xi$ of the boundary data $[\varkappa_a,l^a,m_a]$ that are generated by a vertical diffeomorphism of $(\varkappa_a,l^a,m_a)$, but preserve all other configuration variables, e.g.\ $\delta^{\mtext{diffeo}}_\xi[l^a]=[\xi,l^a]$, but $\delta^{\mtext{diffeo}}_\xi[e_{AA'}]=0$.

The same statement also holds for the shift symmetry, which is generated by the  field variation,
\begin{equation}
\delta^{\mtext{shift}}_\xi[\varkappa_a]=\bar\zeta m_a+\zeta \bar{m}_a\label{shiftdef}
\end{equation}
that annihilates all other bulk and boundary configuration variables other than $\varkappa_a$, e.g.\ $\delta^{\mtext{shift}}_\xi[e_{AA'a}]=0$. The variation of the action \eref{actndef} under any such shifts is given by
\begin{equation}
\delta^{\mtext{shift}}_\xi[S] = -\frac{\I}{16\pi\gamma G}\Big[(\gamma+\I)\int_{\mathcal{N}}\eta_A\ell^A\wedge(\zeta\bar{m}+\bar{\zeta}m)\Big]+\CC
\end{equation}
If the equations of motion are satisfied, the $SL(2,\C)$-invariant scalar $\eta_{Aab}\ell^A$ turns into the area two-form $\varepsilon_{ab}=-2\I m_{[a}\bar{m}_{b]}$. Since $m\wedge\bar{m}\wedge m=0$, the action is trivially stationary with respect to any such variations, $\delta^{\mtext{shift}}_\xi[S]\big|_{\mathrm{EOM}}=0$.

Next, we consider the dilations \eref{dilatdef1}, which generate the infinitesimal field variation
\begin{align}
\delta_f^{\mtext{dilat}}[\varkappa_a]=\partial_a f,\label{dilat1}\\
\delta_f^{\mtext{dilat}}[l^a]=fl^a.\label{dilat2}
\end{align}
All other bulk and boundary configuration variables are annihilated by $\delta_f^{\mtext{dilat}}[\cdot]$. The resulting derivative of the action is
\begin{align}\nonumber
\delta_f^{\mtext{dilat}}[S]\Big|_{\mathrm{EOM}} & = \frac{\I}{8\pi\gamma G}(\gamma+\I)\int_{\mathcal{N}}\Big(-\frac{1}{2}\eta_A\ell^A\wedge \di f+ fN^A\wedge l\hook\eta_A\Big)\Big|_{\mathrm{EOM}} +\CC =\\
& = \frac{\I}{8\pi\gamma G}(\gamma+\I)\int_{\mathcal{N}}\Big(\frac{\I}{2}\varepsilon\wedge \di f+ fN^A\wedge l\hook\eta_A\Big)\Big|_{\mathrm{EOM}} +\CC,\label{dilatvar}
\end{align}
were we used the gluing condition \eref{glucond} to go from the first to the second line. If \eref{bndryeq1} and \eref{bndryeq2} are satisfied, it follows that \eref{dilatvar} is a total exterior derivative. If we compute the expansion of the null surface, we obtain, in fact
\begin{align}\nonumber
-\I\,\di\wedge\varepsilon & = ( D\wedge \eta_A)\ell^A+\eta_A\wedge D\ell^A=-N_A\wedge \bar{m}\ell^A+\eta_A\wedge l\hook N^A=\\
& =-N^A\wedge l\hook \eta_A+\eta_A\wedge l\hook N^A = -2\, N^A\wedge l\hook \eta_A.
\end{align}
Inserting this expression back into \eref{dilatvar}, we obtain a total exterior derivative,
\begin{equation}
\delta_f^{\mtext{dilat}}[S]\Big|_{\mathrm{EOM}} = -\frac{1}{16\pi \gamma G}\Big[(\gamma+\I)\int_{\mathcal{N}}\di\wedge(f\varepsilon)\Big]+\CC=-\frac{1}{8\pi G}\oint_{\partial\mathcal{N}}f\,\varepsilon.
\end{equation}

Finally, we also need to consider the action of the complexified conformal transformations \eref{confdef1}. We decompose the gauge parameter $\lambda$ into its real and imaginary parts \eref{munudef}, $\lambda=\mu+\I\nu$, and obtain the field variation
\begin{align}
\delta_\lambda^{\mtext{conf}}[\varkappa_a]&=-\frac{1}{\gamma}\partial_a\nu,\\
\delta_\lambda^{\mtext{conf}}[l^a]&=\mu\,l^a,\\
\delta_\lambda^{\mtext{conf}}[m_a]&=(\mu+\I\nu)m_a.
\end{align}
 A short calculation yields
\begin{align}\nonumber
\delta_\lambda^{\mtext{conf}}[S]\Big|_{\mathrm{EOM}} &=\frac{\I}{8\pi\gamma G}(\gamma+\I)\int_{\mathcal{N}}\Big[\frac{1}{2\gamma}\eta_A\ell^A\wedge\di \nu-\I\mu\,\omega\wedge m\wedge\bar{m}-\I\nu\, N^A\ell_A\wedge\bar{m}\Big]\Big|_{\mathrm{EOM}}+\CC=\\
&=\frac{\I}{8\pi\gamma G}(\gamma+\I)\int_{\mathcal{N}}\Big[-\frac{\I}{2\gamma}\varepsilon\wedge\di \nu+\mu\,\omega\wedge\varepsilon+\I\nu\, N^A\wedge l\hook\eta_A\Big]\Big|_{\mathrm{EOM}}+\CC\nonumber=\\
&=\frac{\I}{8\pi\gamma G}(\gamma+\I)\int_{\mathcal{N}}\Big[-\frac{\I}{2\gamma}\varepsilon\wedge\di \nu+\frac{\gamma-\I}{\gamma}\mu\,\Gamma\wedge\varepsilon-\frac{1}{2}\nu\, \di\wedge\varepsilon\Big]\Big|_{\mathrm{EOM}}+\CC
\end{align}
Going from the first to the second line, we used the decomposition of $\eta_{Aab}$ wit respect to $(k_a,m_a,\bar{m}_a)$, as given in \eref{etadef}. In the last step, we used the reality cognition \eref{Gammadef} for the abelian connection $\Gamma_a$. Since $\Gamma_a$ is real, we are left with a total derivative that turns into a surface integral,
\begin{align}\nonumber
\delta_\lambda^{\mtext{conf}}[S]\Big|_{\mathrm{EOM}} =\frac{1}{8\pi\gamma G}\int_{\mathcal{N}}\Big[\varepsilon\wedge\di \nu+\nu\, \di\wedge\varepsilon\Big]=\frac{1}{8\pi\gamma G}\oint_{\partial\mathcal{N}}\nu\,\varepsilon.
\end{align}

Let us thus summarise: at its saddle points, the coupled bulk plus boundary action \eref{actndef} is invariant under those variations of the boundary data $(\varkappa_a,l^a,m_a)$  that are generated by vertical diffeomorphisms $\delta_\xi^{\mtext{diffeo}}[\cdot]$, shifts $\delta_\zeta^{\mtext{shift}}[\cdot]$ of $\varkappa_a$, complexified conformal transformations $\delta_{\mu+\I\nu}^{\mtext{dilat}}$ and dilations $\delta_f^{\mtext{dilat}}[\cdot]$. The boundary conditions for the gauge parameters  $(\xi^a,\nu,f)$ are $\xi^a\in T\mathcal{N}:\pi_\ast\xi^a =0, \xi^a\big|_{\partial\mathcal{N}}=0$ and $\nu\big|_{\partial\mathcal{N}}=\gamma\, f\big|_{\partial\mathcal{N}}$, see \eref{diff0def} and \eref{nubndry}. The residual degrees of freedom, which are given by the gauge equivalence class \eref{gravitondef}, are the two radiative modes that pass through the null boundary. Notice that the definition of the equivalence $[\varkappa_a,l^a,m_a]/_\sim$ makes no reference to the interior of spacetime. All fields are intrinsic to the abstract null surface boundary $\mathcal{N}\subset P(\pi,\mathcal{C})$.

\subsection{Radiative symplectic structure}
\noindent The null boundary $\mathcal{N}$ is equipped with a universal ruling that is shared between different spacetimes (points on phase space) such that the fibres $\gamma_z=\pi^{-1}(z), z\in\mathcal{C}$ that generate the null surface are the same for all configurations on the space of (kinematical) histories $\mathcal{H}_{\mtext{kin}}$, see \hyperref[tab1]{table 1} for a summary. If $\bbvar{d}$ is the differential on the space of kinematical histories, we thus have $\bbvar{d}[l^a]\propto l^a$ and $0=\bbvar{d}[l^a]m_a=-l^a\bbvar{d}[m_a]$, where $m_a\in\Omega^1(\mathcal{N}:\C)$ is the co-dyad on $\mathcal{N}$. In other words,
\begin{equation}
\bbvar{d}m_a = \bbvar{f}\,m_a + \bbvar{h}\,\bar{m}_a,\label{mvar}
\end{equation}
where $\bbvar{f}$ and $\bbvar{h}$ are one-forms on field space. The pre-symplectic potential $\Theta_{\mathcal{N}}$ along $\mathcal{N}$ is obtained from the variation of the action and the pull-back on field space to the space of physical histories. Since $\mathcal{N}$ has a boundary, there is a non-trivial cohomology and the pre-symplectic potential on $\mathcal{N}$ is unique only up to the addition of two-dimensional corner terms. In fact, the pre-symplectic potential $\Theta_{\mathcal{N}}$ is the integral of a symplectic current $j_{\mathcal{N}}$, which is a one-form on field space and a three-form on $\mathcal{N}$. Given an action, the pre-symplectic current is unique only up to the addition of an exact one-form, i.e.\ $\tilde{\jmath}_{\mathcal{N}}=j_{\mathcal{N}}+\di{\alpha}$, where $\alpha$ is a one-form on field space and a two-form along $\mathcal{N}$. Reference \cite{DePaoli:2018erh,Barbero2021qiz} provide a more detailed discussions of such corner ambiguities in general relativity. Given the action \eref{actndef}, a possible choice for the pre-symplectic potential along the null surface boundary is given by the one-form on field space
\begin{align}\nonumber
\Theta_{\mathcal{N}}&=\frac{\I}{8\pi\gamma G}(\gamma+\I)\int_{\mathcal{N}}\Big(-\frac{1}{2}\eta_A\ell^A\wedge\bbvar{d}\varkappa-k_a\bbvar{d}l^a\,N^A\wedge l\hook\eta_A +N^A\ell_A\wedge \bbvar{d}\bar{m}\Big)+\CC=\\
&= \frac{\I}{8\pi\gamma G}(\gamma+\I)\int_{\mathcal{N}}\Big(\frac{\I}{2}\varepsilon\wedge\bbvar{d}\varkappa+\frac{\I}{2}k_a\bbvar{d}l^a\,\di\wedge\varepsilon -k\wedge\ell_AD\ell^A\wedge\bbvar{d}\bar{m}\Big)+\CC,\label{thetadef1}
\end{align}
where we used the boundary field equations \eref{glucond} and \eref{bndryeq1} to go from the first to the second line. 

Since the four-dimensional region $\mathcal{M}$ is also bounded by the spacelike disks $M_0$ and $M_1$, we  also have corresponding contributions to the pre-symplectic potential, 
\begin{equation}
\Theta_{M}= \frac{\I}{8\pi\gamma G}(\gamma+\I)\left[\int_{M}\Sigma_{AB}\wedge\bbvar{d}A^{AB}-\oint_{\partial M}\eta_A\bbvar{d}\ell^A\right]+\CC
\end{equation}
If $\delta$ denotes a linearised solution of the bulk plus boundary field theory (a tangent vector to $\mathcal{H}_{\mtext{phys}}$), we obtain
\begin{equation}
\delta[S]\big|_{\mathcal{H}_{\mtext{phys}}}=\Theta_{\mathcal{N}}(\delta)+\Theta_{M_1}(\delta)-\Theta_{M_0}(\delta).
\end{equation}

Since we are primarily interested in the symplectic structure along the null boundary, let us further understand the nature of the various terms that appear in the pre-symplectic potential \eref{thetadef1}. Since $l^a\ell_AD_a\ell^A=0$, which follows from \eref{bndryeq1},  the one-form $\ell_AD\ell^A$ on $\mathcal{N}$ admits the decomposition
\begin{equation}
\ell_AD\ell^A = -\left(\frac{1}{2}\vartheta_{(l)}{m}+{\sigma}_{(l)}\bar{m}\right).\label{lldef}
\end{equation}
The components $\sigma_{(l)}$ and $\vartheta_{(l)}$ are the shear and expansion of the null congruence $l^a\in T\mathcal{N}$. Going back to \eref{tetranull}, we obtain, in fact
\begin{align}
\ell_Am^aD_a\ell^A & = \bar{k}_{A'}\ell_Am^aD_a(\bar{\ell}^{A'}\ell^A)=-{m}_{b}m^aD_al^b\Neq-\hat{m}^a\hat{m}^b\nabla_al_b=-\sigma_{(l)},\label{sheardef}\\
\ell_A\bar{m}^aD_a\ell^A & = \bar{k}_{A'}\ell_A\bar{m}^aD_a(\bar{\ell}^{A'}\ell^A)=-{m}_{b}\bar{m}^aD_al^b\Neq-\hat{q}^{ab}\nabla_a l_b=-2\vartheta_{(l)}\label{expansdef},
\end{align}
where we extended in the last step $l^a\in T\mathcal{N}$ and $m^a\in T_\C\mathcal{N}$ into vector fields $\hat{l}^a\in T\mathcal{M}$ and $\hat{m}^a\in T_\C\mathcal{M}$, which are defined in a neighbourhood of $\mathcal{N}$, and $\hat{q}^{ab}=2 \hat{m}^{(a}\hat{\bar{m}}^{b)}$. This step is necessary to identify the components of $\ell_A D\ell^A$ with shear and expansion of the null congruence $\hat{l}^a\big|_{\mathcal{N}}\in T\mathcal{M}$. The symbol $\Neq$ stands for \emph{equals on $\mathcal{N}$}. It is important to note, however, that it is possible to compute shear and expansion directly from the exterior derivative of the boundary intrinsic one-form $m_a\in\Omega^1(\mathcal{N}:\C)$ alone. An embedding of the boundary into the bulk is not necessary to infer shear and expansion. In fact, the exterior derivative of the one-form $m_a$ on $\mathcal{N}$ admits the expansion
\begin{equation}
\di\wedge m = -\I \big(\varphi_{(l)}k+\gamma \bar{m}\big)\wedge m - \frac{1}{2}\vartheta_{(l)} k\wedge m - \sigma_{(l)}k\wedge \bar{m}.\label{dim}
\end{equation}
Notice that the one-form $\varphi_{(l)}k+\gamma m +\bar{\gamma}m$, defines a $U(1)$ spin connection on $\mathcal{N}$. The definition of this connection depends on a choice for the one-form $k_a:k_al^a=-2$. If we replace $k_a$ by $k_a+f m_a+\bar{f} m_a$, he spin coefficient $\gamma$ is shifted into $\gamma- f\varphi_{(l)}+\frac{\I}{2}\vartheta_{(l)}f-\I\sigma_{(l)}\bar{f}$. On the other hand, the spin coefficient $\varphi_{(l)}$ equals $l^a\Gamma_a$, with $\Gamma_a$ denoting the $U(1)$ connection introduced in \eref{Gammadef}. Using the spinor formalism, the proof is immediate: 
\begin{align}\nonumber
2\I\varphi_{(l)}&\Neq2\hat{l}^a\hat{\bar{m}}^b\nabla_{[a}\hat{m}_{b]}-\CC  
\Neq \hat{l}^a\hat{\bar{m}}^b\nabla_a \hat{m}_b +\hat{\bar{m}}^a\hat{m}^b\nabla_a \hat{l}_b-\CC\Neq\\
\nonumber & \Neq -\left(k_A\bar{\ell}_Al^a D_a (\ell^A\bar{k}^{A'})-\CC\right)+2\hat{\bar{m}}^a\hat{m}^b\nabla_{[a}\hat{\ell}_{b]} \Neq\\
&\Neq2 k_Al^a D_a\ell^A-\CC = 2\I\ell^a\Gamma_a.\label{phidef}
\end{align}
where we used the Frobenius integrability condition $\hat{l}_{[a}\nabla_b \hat{l}_{c]}\Neq0$ to eliminate the second term in the second line. In addition, we used the boundary equations of motion \eref{bndryeq1}, \eref{Gammadef} and the isomorphism \eref{tetranull} between spinors and vectors on the null boundary to arrive at the final result.

Another useful equation is to write the expansion $\vartheta_{(l)}$ of the null surface solely in terms of the exterior derivative of the area two-form $\varepsilon\in\Omega^2(\mathcal{N}:\R)$,
\begin{equation}
\di\wedge\varepsilon = -\vartheta_{(l)}\, k\wedge\varepsilon\label{expansdef2}
\end{equation}

Taking into account $k_a\bbvar{d}l^a=-l^a\bbvar{d}{k}_a$, we can then write the pre-symplectic potential \eref{thetadef1} in the following simplified form
\begin{align}
\Theta_{\mathcal{N}}&=
-\frac{1}{8\pi G}\int_{\mathcal{N}}\varepsilon\wedge\bbvar{d}\varkappa+\frac{\I}{8\pi\gamma G}\int_{\mathcal{N}}\left((\gamma+\I)\ell_AD\ell^A\wedge\bbvar{d}(k\wedge \bar{m})-\CC\right).\label{thetadef2}
\end{align}
The corresponding pre-symplectic two-form is given by the exterior derivative on field space
\begin{equation}
\Omega_{\mathcal{N}}=\bbvar{d}\Theta_{\mathcal{N}}.\label{Omdef}
\end{equation}
\subsection{Gauge symmetries on the radiative phase space}
\noindent In this section, we identify the gauge symmetries of the pre-symplectic two-form \eref{Omdef}, which are the degenerate directions of $\Omega_{\mathcal{N}}$. These are given by combinations of (i) dilatations of the null normal $l^a$, (ii) $U(1)$ transformations of the boundary data $(\varkappa_a,m_a)$, (iii) shifts of $\varkappa_a$, and (iv) horizontal diffeomorphisms of $(\varkappa_a,l^a,m_a)$.\medskip 

\emph{(i) dilatations:} First of all, we consider the dilatations \eref{dilat1}, \eref{dilat2} of the boundary data. An infinitesimal such variation defines the vector field $\delta_f^{\mtext{dilat}}[\cdot]$, whose components are given by \eref{dilat1} and \eref{dilat2}. Going back to the definition of shear and expansion, \eref{sheardef}, \eref{expansdef}, \eref{lldef}, we obtain

\begin{equation}
\delta_f^{\mtext{dilat}}[\ell_AD\ell^A]=f\ell_AD\ell^A.
\end{equation}
Since $k_al^a=-1$, and $\delta_f^{\mtext{dilat}}[l^a]=fl^a$, we then also know
\begin{equation}
\ell_A D\ell^A\wedge \delta_f^{\mtext{dilat}}[k\wedge\bar{m}]=-f \ell_A D\ell^A\wedge k\wedge\bar{m}=-\frac{1}{2}\,f\,\vartheta_{(l)}\,k\wedge m\wedge\bar{m}.
\end{equation}
Therefore,
\begin{align}\nonumber
\Omega_{\mathcal{N}}(\delta_f^{\mtext{dilat}.},\delta) &=-\frac{1}{8\pi G}\int_{\mathcal{N}}\delta\varepsilon\wedge\di f-\frac{\I}{16\pi\gamma G}\int_{\mathcal{N}}\left((\gamma+\I)f\,\vartheta_{(l)}\,k\wedge m\wedge\bar{m}-\CC\right)=\\
&= \frac{1}{8\pi G}
\int_{\mathcal{N}}\left(\delta\varepsilon\wedge\di f + f\di\wedge\delta\varepsilon\right)= -\frac{1}{8\pi G}\oint_{\partial\mathcal{N}}f\,\delta\varepsilon=-\delta\big(K_f[\mathcal{C}_1]-K_f[\mathcal{C}_0]\big).\label{Kzero}
\end{align}
Where we defined the corresponding charge,
\begin{equation}
K_f[\mathcal{C}]=\frac{1}{8\pi G}\oint_{\mathcal{C}}f\,\varepsilon.\label{Kdef}
\end{equation}
The field variation $\delta_f^{\mtext{dilat}}[\cdot]$ defines an unphysical gauge direction, if the  gauge parameters $f:\mathcal{N}\rightarrow\R$ has compact support. A large gauge transformation, where $f$ does not vanish at the boundary, is generated by the dilatation charge given by \eref{Kdef}.\medskip

\emph{(ii) $U(1)$ transformations:} Next, there are the $U(1)$ transformations,
\begin{align}
\delta^{U(1)}_\nu[\varkappa_a]&=-\gamma^{-1}\partial_a\nu,\label{U1kappa}\\
\delta^{U(1)}_\nu[l^a]&=0,\label{U1l}\\
\delta^{U(1)}_\nu[m_a]&=\I\nu\,m_a.\label{U1m}
\end{align}
Notice that $\gamma\kappa_a$ transforms as a $U(1)$ connection. On the other hand, the null vector $l^a$ is charge neutral and the one-form $m_a$ is a vector under $U(1)$.

Given \eref{U1m}, \eref{lldef}, \eref{sheardef} and \eref{expansdef}, we then also know
\begin{equation}
\delta_\nu^{U(1)}[\ell_AD\ell^A]=\I\nu\,\ell_AD\ell^A.
\end{equation}
Bringing everything together, and taking into account that $\varepsilon=-\I m\wedge\bar{m}$, we obtain
\begin{align}\nonumber
\Omega_{\mathcal{N}}(\delta_{\nu}^{U(1)},\delta)&=-\frac{1}{8\pi\gamma G}\int_{\mathcal{N}}\di\nu\wedge\delta\varepsilon+\frac{1}{8\pi\gamma G }\int_{\mathcal{N}}\left((\gamma+\I)\nu\,\delta(k\wedge\ell_A D\ell^A\wedge\bar{m})+\CC \right)=\\\nonumber
&=-\frac{1}{8\pi\gamma G}\int_{\mathcal{N}}\di\nu\wedge\delta\varepsilon-\frac{1}{16\pi\gamma G}\int_{\mathcal{N}}\left((\gamma+\I)\nu\,\delta\left(\vartheta_{(l)}k\wedge m\wedge\bar{m}\right)+\CC\right)=\\\nonumber
&=-\frac{1}{8\pi\gamma G}\int_{\mathcal{N}}\di\nu\wedge\delta\varepsilon-\frac{1}{8\pi\gamma G}\int_{\mathcal{N}}\nu\,\di\wedge(\delta\varepsilon)=\\
&=-\delta\big(L_\nu[\mathcal{C}_1]-L_\nu[\mathcal{C}_0]\big).\label{U1gen}
\end{align}
Where we introduced the $U(1)$ charge
\begin{equation}
L_\nu[\mathcal{C}]=\frac{1}{8\pi\gamma G}\oint_{\mathcal{C}}\nu\,\varepsilon.\label{Ldef}
\end{equation}

Going back to \eref{Kdef}, we obtain an important constraint between the dilatation charge \eref{Kdef} and the $U(1)$ charge \eref{Ldef}, namely
\begin{equation}
K_{f}[\mathcal{C}] = \gamma L_f[\mathcal{C}].\label{KLcons}
\end{equation}
 The orbits of the $U(1)$ charge \eref{Ldef} are compact. Since the orbits are compact, the corresponding charge
 must be quantised \cite{PhysRevD.1.2360}. On the other hand, the $U(1)$ generator $L_f[\mathcal{C}]$ is  the surface integral of the area density multiplied by the function $f:\mathcal{C}\rightarrow \R$. The quantisation of the $U(1)$ charge implies, therefore, the quantisation of area in units of $\gamma\ellp^2$, with $\ellp^2=8\pi\hbar G/c^3$ denoting the Planck area. This observation provides a simple explanation for the quantisation of area in loop quantum gravity \cite{Rovelliarea,PhysRevD.47.1703,AshtekarLewandowskiArea,Immirziparam,Barbero1994,Wieland:2017cmf}.\medskip
 
\emph{(iii) Shifts of $\varkappa_a$:} under the shift symmetry, the one-form $\varkappa_a$ transforms according to \eref{shiftdef}. All other boundary fields $(l^a,m_a,\bar{m}_a)$ are annihilated by the action of $\delta_\zeta^{\mtext{shift}}$. We thus have
\begin{align}
\Omega_{\mathcal{N}}(\delta_\zeta^{\mtext{shift}},\delta) = +\frac{1}{8\pi G}\int_{\mathcal{N}}\delta\varepsilon\wedge(\zeta\bar{m}+\bar{\zeta}m).
\end{align}
On the other hand, $\varepsilon=-\I m \wedge\bar{m}$. Taking into account the variation of the co-dyad \eref{mvar}, we obtain
\begin{align}
\Omega_{\mathcal{N}}(\delta_\zeta^{\mtext{shift}},\delta) = 0,\label{shiftzero}
\end{align}
such that the shift symmetry defines a degenerate null direction of the pre-symplectic two-form. \medskip

\emph{(iv) vertical diffeomorphisms} Finally, we return to the vertical diffeomorphisms, which are generated by vertical vector fields $\xi^a\in T\mathcal{N}:\pi_\ast\xi^a=0$ that vanish at the boundary of $\mathcal{N}$, i.e.
\begin{equation}
\xi^a\big|_{\partial\mathcal{N}}=0.
\end{equation}
The resulting infinitesimal field variation is the Lie derivative
\begin{align}
\mathcal{L}_\xi \varkappa & = \xi\hook(\di\wedge\varkappa)+\di\wedge(\xi\hook \varkappa),\\
\mathcal{L}_\xi l^a & = [\xi,l]^a,\\
\mathcal{L}_\xi m & = \xi\hook(\di\wedge m),
\end{align}
where $[\cdot,\cdot]$ is the Lie bracket of vector fields on $\mathcal{N}$ and $\xi\hook$ denotes the interior product between a vector field and a $p$-form. If such a  vector field $\xi^a$ is field independent, we have $\delta[\xi^a]=0$, and obtain
\begin{align}\nonumber
\Omega_{\mathcal{N}}(\mathcal{L}_\xi,\delta)=&-\frac{1}{8\pi G}\int_{\mathcal{N}}\Big[\mathcal{L}_\xi\varepsilon\wedge\delta\varkappa-\delta\varepsilon\wedge\mathcal{L}_\xi\varkappa\Big]+\\
\nonumber
&\quad+\frac{\I}{8\pi\gamma G}\left[(\gamma+\I)\int_{\mathcal{N}}\Big(\mathcal{L}_\xi(\ell_A D\ell^A)\wedge\delta (k\wedge \bar{m})-\delta(\ell_AD\ell^A)\wedge{\mathcal{L}_\xi}(k\wedge\bar{m})\Big)-\CC\right]=\\\nonumber
= &\,\frac{1}{8\pi G}\int_{\partial{\mathcal{N}}}\xi\hook(\varkappa\wedge\delta\varepsilon)+\\
&\quad-\frac{\I}{8\pi\gamma G}\left[(\gamma+\I)\int_{\partial\mathcal{N}}\Big(\xi\hook\big(\delta(\ell_AD\ell^A)\wedge k\wedge\bar{m}\big)\Big)-\CC\right]-\delta[C_\xi].\label{Cxi}
\end{align}
The two boundary terms disappear, because $\xi^a$ vanishes at $\partial\mathcal{N}$, see \eref{diff0def}. We are thus left with the last term, which is given by the integral
\begin{align}
C_\xi&=\frac{1}{8\pi G}\int_{\mathcal{N}}\mathcal{L}_\xi\varepsilon\wedge\varkappa-\frac{\I}{8\pi\gamma G}\left[(\gamma+\I)\int_{\mathcal{N}}\big(\mathcal{L}_\xi(\ell_AD\ell^A)\wedge k\wedge\bar{m}\big)-\CC\right].\label{Cxi1}
\end{align}
This integral vanishes as a constraint. In fact, taking into account that $\xi\hook(\ell_AD\ell^A)=0$, see \eref{bndryeq1}, we obtain
\begin{align}\nonumber
\mathcal{L}_\xi(\ell_AD\ell^A)\wedge k\wedge \bar{m}&=\left(\xi\hook (D\ell_A\wedge D\ell^A)+\xi\hook(\ell_A D\wedge D \ell^A)\right)\wedge k\wedge \bar{m}=\\
\nonumber&=\left(2(\xi\hook D\ell_A)D\ell^A-\xi\hook F_{AB}\ell^A\ell^B\right)\wedge k\wedge \bar{m}=\\
\nonumber &=\left(\xi\hook(\varkappa+\I\omega)\ell_AD\ell^A-\xi\hook F_{AB}\ell^A\ell^B\right)\wedge k\wedge \bar{m}=\\
 &=\frac{1}{2}\xi\hook(\varkappa+\I\omega)\vartheta_{(l)}k\wedge m\wedge\bar{m}-(\xi\hook F_{AB})\ell^A\ell^B \wedge k\wedge \bar{m}.\label{Cxi2}
\end{align}
The first term on the right hand side is proportional to the expansion of the null congruence, which follows directly from \eref{expansdef2}. The second term can be simplified by taking  into account that the pull back of the tetrad $e_{AA'}$ to the null surface satisfies the gluing condition
\begin{equation}
\varphi^\ast_{\mathcal{N}}e_{AA'}\bar{\ell}^{A'}=\I\ell_A\bar{m},\label{Cxi3}
\end{equation}
where the fields on the right hand side of this equation are all intrinsic to the null surface. Inserting \eref{Cxi3} back into to \eref{Cxi2} and taking into account he reality condition \eref{Gammadef} for the one-form $\omega_a$, the generator $C_\xi$ becomes
\begin{align}\nonumber
C_\xi = \frac{1}{8\pi G}\int_{\mathcal{N}}&\Big(\xi\hook(\di\wedge\varepsilon)\wedge\varkappa-(\xi\hook\varkappa)\wedge\di\wedge\varepsilon\big)+\\
&+\frac{1}{8\pi\gamma G}\left[(\gamma+\I)\int_{\mathcal{N}}(\xi\hook F_{AB})\wedge \ou{e}{B}{A'}\ell^A\bar{\ell}^{A'}\wedge k-\CC\right].
\end{align}
The first two terms on the right hand side cancel each other. The third term vanishes as well---provided the Einstein equations \eref{EEQ} are satisfied. If we insert, in fact, the Einstein equations in addition to $\xi\hook e_{AA'}=\xi_{AA'}=-\I\xi\hook k\,\ell_A\bar{\ell}_{A'}$ and $\ell_A\ell^A=0$, we obtain
\begin{align}\nonumber
C_\xi &= -\frac{\I}{8\pi\gamma G}\left[(\gamma+\I)\int_{\mathcal{N}} F_{AB}\wedge\ou{e}{B}{A'}\xi^{AA'}-\CC\right]=\\
&=-\frac{\I\Lambda}{24\pi\gamma G}\left[(\gamma+\I)\int_{\mathcal{N}} \Sigma_{AB}\wedge\ou{e}{B}{A'}\xi^{AA'}-\CC\right]=0.\label{Cxi4}
\end{align}
The last line vanishes: the vector-valued volume three-form is given by
\begin{equation}
d^3v_{AA'}=\frac{2\I}{3}\Sigma_{AB}\wedge \ou{e}{B}{A'}.
\end{equation}
Its pull-back to the null surface is proportional to the null normal. In fact,
\begin{equation}
\varphi^\ast_{\mathcal{N}}d^3v_{AA'}=-\ell_A\bar{\ell}_{A'}k\wedge m\wedge \bar{m}.
\end{equation}
Since $\xi^a$ is a vertical vector field, it satisfies $\xi^ae_{AA'a}\propto\I\ell_A\bar{\ell}_{A'}$, such that $\xi^{AA'}\varphi^\ast_{\mathcal{N}}d^3v_{AA'}=0$. Going back to \eref{Cxi4}, we thus see that the generator $C_\xi$ of vertical diffeomorphisms \eref{diffeodef1} vanishes as a constraint.   Therefore,
\begin{equation}
\forall \xi^a\sim l^a, \xi^a\big|_{\partial\mathcal{N}}=0 : \Omega_{\mathcal{N}}(\mathcal{L}_\xi,\delta)=-\delta C_\xi =0.\label{diffeozero}
\end{equation} We will see in the next section that the constraint $C_\xi=0$ is nothing but the Raychaudhuri equation for the null congruence $l^a\in T\mathcal{N}$.

\medskip

\emph{Summary:} Before we go on and proceed to the second half of the paper, let us briefly summarise. The starting point of this section was the definition of the bulk plus boundary action with the parity odd Holst term in the bulk and a boundary that is null. The null boundary is homeomorphic to the cartesian product $\R\times\mathcal{C}$ and it is equipped with a universal structure $P(\mathcal{C},\pi)$, which consists of a surjection $\pi:P\rightarrow\mathcal{C}$ onto the base manifold $\mathcal{C}$ that determines the ruling of the null boundary by its null rays $\gamma_z=\pi^{-1}(z)\subset P, z\in\mathcal{C} $. The boundary conditions are such that a gauge equivalence class of boundary fields $[\varkappa_a,l^a,m_a]\in T^\ast\mathcal{N}\times T\mathcal{N}\times T^\ast_{\C}\mathcal{N}$ is kept fixed. The one-form $\varkappa_a$ encodes the non affinity of the null vector field $l^a\in T\mathcal{N}$. The complex-valued one-form $m_a$ determines the degenerate metric $q_{ab}=2m_{(a}\bar{m}_{b)}$ on the null boundary. The inner product of $m_a$ and $l^a$ vanishes, i.e.\ $m_al^a=0$. The gauge equivalence class $[\varkappa_a,l^a,m_a]/_\sim$ describes two physical degrees of freedom along the null boundary, which are the two degrees of freedom of gravitational radiation crossing the null boundary.\footnote{The counting is immediate: since $l^a\in T\mathcal{N}$ lies parallel to the null rays, it is characterised by a single number (a lapse function). Since $l^am_a=0$, see \eref{lmzero}, there are two complex components left to determine $m_a\in T^\ast\mathcal{N}$. Finally, there is $\varkappa_a$, which is a real-valued one form intrinsic to $\mathcal{N}$. Hence there are $1+2\times 2+3=8$ local degrees of freedom. The vertical diffeomorphisms \eref{diff0def} remove one of them, the shift symmetry \eref{shiftdef1} another two, the dilatations \eref{dilatdef1} remove the overall scale of $l^a$ and the complexified conformal transformations  \eref{confdef1} remove two degrees of freedom as well. The gauge equivalence class $\mathcal{g}=[\varkappa_a,l^a,m_a]/_\sim$ dexcibrs therefore two degrees of freedom along $\mathcal{N}$.} Given the parity-odd action in the bulk, we introduced the resulting pre-symplectic two-form along the null boundary and studied its gauge symmetries, which consist of  dilatations of $(\varkappa_a,l^a)$, $U(1)$ gauge transformations of $(\varkappa_a,m_a)$, shifts of $\varkappa_a$ and vertical diffeomrophisms, see \eref{Kzero}, \eref{U1gen}, \eref{shiftzero}, \eref{diffeozero}.   Notice that $\varkappa_a$ is an abelian connection with resect to  both $U(1)$ gauge transformations  as well as dilations, see \eref{dilat1} and \eref{U1kappa}. This is an important observation and it is crucial to understand the quantisation of area in loop quantum gravity \cite{AshtekarLewandowskiArea,Rovelliarea,PhysRevD.47.1703,Immirziparam,Barbero1994,Wieland:2017cmf}.

\section[$\boldsymbol{SL(2,\R)}$ boundary charges]{$SL(2,\R)$ boundary charges}\label{sec3}
\noindent In this section, we derive the Poisson commutation relations among the  configuration variables on the null cone. The construction has three steps. The first step is to identify an implicit  $SL(2,\R)$-symmetry among the boundary fields $[\varkappa_a,l^a,m_a]$. The next step is to embed the covariant phase space into an auxiliary (kinematical) phase space, where the $SL(2,\R)$ symmetry is manifest. The final step is to impose the constraints and calculate the Dirac bracket to identify the Poisson commutation relations among the physical observables.
\subsection{$SL(2,\R)$ parametrisation of the boundary fields}
\noindent To parametrise the boundary fields $[\varkappa_a,l^a,m_a]$, it is useful to work with auxiliary $SL(2,\R)$ variables on the null boundary.  Over every point of $\mathcal{N}$, we introduce an internal and two-dimensional auxiliary vector space $\mathbb{V}$ over $\R$, which will carry a natural representation of $SL(2,\R)$. We equipp this vector space with a complex structure $J$ and a compatible metric $q$. Let then $\{\boldsymbol{f}_i, i=1,2\}$ be an orthonormal basis of $(\mathbb{V},J,q)$ such that
\begin{align}
q(\boldsymbol{f}_i,\boldsymbol{f}_j) &\equiv q_{ij} = \delta_{ij},\label{intrnlq}\\
J(\boldsymbol{f}_1) & = + \boldsymbol{f}_2,\\
J(\boldsymbol{f}_2) & = - \boldsymbol{f}_1,
\end{align}
where $\delta_{ij}$ is the Kronecker delta. Given the complex co-dyad $(m,\bar{m})$ on $\mathcal{N}$, we introduce the associated $\mathbb{V}$-valued differential form, 
\begin{equation}
\boldsymbol{e} = \frac{1}{\sqrt{2}}\left(m+\bar{m}\right)\otimes\boldsymbol{f}_1 -\frac{\I}{\sqrt{2}}\left(m-\bar{m}\right)\otimes\boldsymbol{f}_2= e^i\otimes\boldsymbol{f}_i,
\end{equation}
Next, we decompose the one-form $e^i$ with respect to the original complex-valued dyad $(m_a,\bar{m}_a)$,
\begin{equation}
e^i = m^i\bar{m}+\bar{m}^im,\label{edecomp}
\end{equation}
where $m^i$ ($\bar{m}^i$) are the components of $\ou{e}{i}{a}$ with respect to the co-vector $\bar{m}_a$ ($m_a$). In the following, we will use the internal metric $q_{ij}$ and its inverse $q^{ij}:q^{im}q_{mj}=\delta^i_j$ to raise and lower vector indices in $\mathbb{V}$, which is two-dimensional. Accordingly,
\begin{equation}
m_i = q_{ij}m^j.
\end{equation}
We then also have the internal volume two-form,
\begin{equation}
\varepsilon_{ij}  = \ou{J}{m}{j} q_{mi}= -2\I m_{[i}\bar{m}_{j]},
\end{equation}
where $\ou{J}{m}{i}$ are the components of the complex structure: $J(\boldsymbol{e}_i)=\ou{J}{m}{i}\boldsymbol{e}_m$.\medskip

The vector space $\mathbb{V}$ is equipped with a natural representation of $SL(2,\R)$. A linear map $S:\mathbb{V}\rightarrow\mathbb{V}$ defines, in fact, an element of $SL(2,\R)$, if it preserves the volume two-form. In components
\begin{equation}
\varepsilon_{lm}\ou{S}{l}{i}\ou{S}{m}{j}=\varepsilon_{ij}.
\end{equation}
Such that the inverse matrix is given by
\begin{equation}
\ou{[S^{-1}]}{i}{j}=\varepsilon^{in}\ou{S}{m}{n}\varepsilon_{jm},
\end{equation}
where $\varepsilon^{im}\varepsilon_{jm}=\delta^i_j$. A basis in the corresponding Lie algebra $\mathfrak{sl}(2,\R)$ is given by
\begin{align}
\ou{J}{i}{k} & = \I\left(\bar{m}^im_k-m^i\bar{m}_k\right),\label{Jgen}\\
\ou{X}{i}{k} & = m^im_k,\label{Xgen}\\
\ou{\bar{X}}{i}{k} &= \bar{m}^i\bar{m}_k,\label{barXgen}
\end{align}
where $J$ is a $U(1)$ generator and $X$ is a complex and traceless $2\times 2$ matrix. We obtain the commutation relations,
\begin{align}
[J,X]&=-2\I X,\\
[X,\bar{X}]&=+\I J,
\end{align}
where we suppressed the vector indices $i,j=1,2$. In the following, we will also need a projector $P:\mathfrak{sl}(2,\R)\rightarrow\mathfrak{sl}(2,\R)$  that annihilates the $U(1)$ generator $J\in\mathfrak{sl}(2,\R)$, but otherwise preserves $X$ and $\bar{X}$. In other words,
\begin{equation}
PJ=0,\quad PX=X.\label{Pdef}
\end{equation}
\medskip

Going back to the decomposition of $\ou{e}{i}{a}$ in terms of the co-dyad $(m_a,\bar{m}_a)$, we now also have an induced action of $SL(2,\R)$ onto the complex-valued one-forms $(m_a,\bar{m}_a)$. We call this action $S\triangleright m_a$, and it is defined via
\begin{equation}
\forall S\in SL(2,\R):\ou{S}{i}{j}e^j = m^i(S\triangleright\bar{m})+\bar{m}^i(S\triangleright m).
\end{equation}
The basic idea is to now parametrise the physical co-dyad $m_a$ in terms of a fiducial co-dyad $m_a^{(0)}$, an $SL(2,\R)$ transformation $S:\mathcal{N}\rightarrow SL(2,\R)$ and an overall conformal factor $\Omega =\E^\mu$,
\begin{align}
m_a  &= \Omega\,(S\triangleright m^{(0)}_a),\label{mparam}\\
\Omega &= \E^\mu.\label{conffactr}
\end{align}
Given the fiducial dyad, the corresponding fiducial volume element is
\begin{equation}
d^2v_o := -\I\, m^{(0)}\wedge\bar{m}^{(0)}.
\end{equation}
If the base manifold $\mathcal{C}$ has the topology of a two sphere, a natural choice for $m_a^{(0)}$ is given by
\begin{equation}
m^{(0)}_a = \frac{\partial_az}{1+|z|^2},
\end{equation}
where $z = \cot\frac{\vartheta}{2}\,\E^{\I\varphi}$ maps the unit sphere onto the complex plane.\medskip 

The radiative symplectic potential \eref{thetadef2} depends not only on $m_a$, but also on $\varkappa_a$ and $l^a$ for which we now introduce a convenient parametrisation as well. Consider first the one-form $\varkappa_a$, which can always be parametrised as follows:
\begin{equation}
\varkappa_a = \partial_a\lambda -\frac{1}{\gamma}\varphi_{(l)}k_a,\label{kappaparam}
\end{equation}
where $\lambda:\mathcal{N}\rightarrow \R$ is a gauge parameter for dilatations, $k_a$ is a one-form on $\mathcal{N}$ such that $k_al^a=-1$ and the $U(1)$ gauge potential $\varphi_{(l)}=l^a\Gamma_a$ is inferred from the Lie derivative
\begin{equation}
\mathcal{L}_l m =\left(\frac{1}{2}\vartheta_{(l)}+\I\varphi_{(l)}\right) m+\sigma_{(l)}\bar{m},
\end{equation}
as in \eref{dim} above. Finally, we also need to choose a parametrisation for the null normal $l^a$. This can be done by introducing a fiducial time coordinate $u:\mathcal{N}\rightarrow[-1,1]$ along $\mathcal{N}$, which increases monotonically along the null generators: i.e.\ for every future pointing null vector $[l']^a\in T\mathcal{N}$: $[l']^a\partial_a u> 0$. In addition, we demand that the fiducial (unphysical) $u$-coordinate satisfies the boundary conditions
\begin{equation}
u\big|_{\mathcal{C}_1} =1,\qquad u\big|_{\mathcal{C}_0} =-1,\label{ubndry}
\end{equation}
 at the boundary of $\mathcal{N}$, which consists of two disjoint cross sections: $\partial\mathcal{N}=\mathcal{C}_0\cup\mathcal{C}_1^{-1}$. Since the vector space of all null vectors on $\mathcal{N}$ is one-dimensional, we can parametrize any future pointing null vector $l^a$ in terms of the fiducial null vector $\partial^a_u$ and an overall scale, e.g.
\begin{equation}
l^a =\E^{\mu+\lambda+\rho}\partial^a_u,\label{lparam}
\end{equation}
where $\mu$ and $\lambda$ are the same as in \eref{kappaparam} and \eref{mparam}, but $\E^\rho$ is arbitrary and defines the lapse function
\begin{equation}
\E^\rho = N_{(l)}.\label{Ndef}
\end{equation}

As far as the covariant phase space is concerned both the time function $u:\mathcal{N}\rightarrow [-1,1]$ as well as the fiducial co-dyad $m_a^{(0)}$ are fixed background structures. Their field variations vanish. In other words,
\begin{equation}
\delta[u]=0,\qquad \delta[\partial^a_u]=0, \qquad \delta[m_a^{(0)}]=0.
\end{equation}
 
\subsection{Choice of time}
\noindent  Above, we introduced a fiducial time coordinate $u:\mathcal{N}\rightarrow [-1,1]$. By introducing a (local) homeomorphism $\text{\sl z}:\mathcal{C}\rightarrow \C$ on the base manifold, which naturally extends via $z=\text{\sl z}\circ \pi$ into a coordinate on $\mathcal{N}$,  we have a fiducial coordinate system $(u,z,\bar{z})$ along the null boundary. The $u$ coordinate is unphysical. The only conditions are that it increase monotonically and satisfy the boundary conditions \eref{ubndry}. Otherwise,  the $u$ coordinate is completely arbitrary. A more physical time coordinate can be introduced as follows. Consider first the total (affine) duration of a null generator of $\mathcal{N}$, as given by the affine length
\begin{equation}
L(z,\bar{z}) = \int_{-1}^1\di u\, N_{(l)}^{-1}(u,z,\bar{z}),
\end{equation}
where $N_{(l)}$ is the lapse function \eref{Ndef}. We may then introduce the new time coordinate
\begin{equation}
U(u,z,\bar{z}) =\frac{2\int_{-1}^u\di u'\,N_{(l)}^{-1}(u',z,\bar{u})-L(z,\bar{z})}{L(z,\bar{z})},
\end{equation}
which satisfies the boundary conditions 
\begin{equation}
U(\pm 1,z,\bar{z}) = \pm 1.
\end{equation}

Let us also consider the corresponding null vector field
\begin{equation}
U^a = \frac{L(z,\bar{z})}{2} \E^{-\mu-\lambda}l^a.\label{Uvecdef}
\end{equation}
Its non-affinity is proportional to the expansion of the null surface. This can be seen as follows. Consider first  $\kappa_{(l)}:l^b\nabla_b l^a =\kappa_{(l)}l^a$, which is given by
\begin{equation}
\kappa_{(l)} = l^a\left(\varkappa_a+\gamma^{-1}\Gamma_a\right)=l^a \partial_a\lambda,
\end{equation}
see \eref{nonaffnty}, \eref{phidef} and \eref{mparam}. Therefore,
\begin{equation}
U^b\nabla_b U^a =\frac{L^2}{4}\E^{-\mu-\lambda}l^b\nabla_b\left(\E^{-\mu-\lambda}l^a\right) = -\frac{1}{2}\left(\Omega^{-2}\frac{\di}{\di U}\Omega^2\right) U^a,\label{Unonaff}
\end{equation}
where $\Omega=\E^\mu$ is the conformal factor \eref{conffactr}. Hence, the non-affinity of $U^a$ is given by its expansion. Derivatives with respect to $U^a$ will be denoted as follows: if $F:\mathcal{N}\rightarrow\R$ is a scalar function on the null surface, its time derivative is denoted by
\begin{equation}
\dot{F}:= U^a\partial_a[F]=:\frac{\di}{\di U} F.
\end{equation}

\subsection{Kinematical phase space}
\noindent Next, we need to express the symplectic potential \eref{thetadef2} in terms of the $SL(2,\R)$ variables introduced above. Consider first the Lie derivative of the vector-valued co-dyad $\boldsymbol{e}_a$ along the null generators, see \eref{edecomp}. Since $\boldsymbol{e}_a$ takes values in a two-dimensional vector space, we first have to explain how the Lie derivative acts onto the basis vectors $\boldsymbol{f}_i, i=1,2$. The simplest possibility is
\begin{equation}
\mathcal{L}_l \boldsymbol{f}_i =0.
\end{equation}
With this choice, we obtain
\begin{equation}
\mathcal{L}_l e^i = \left(\Omega^{-1}\mathcal{L}_l\Omega\right)e^i + \ou{\left[\mathcal{L}_l S\cdot S^{-1}\right]}{i}{j}e^j,
\end{equation}
where
\begin{align}
\Omega^{-1}\mathcal{L}_l\Omega & = \frac{1}{2}\vartheta_{(l)},\label{voldot}\\
\mathcal{L}_l S\cdot S^{-1} & = \varphi_{(l)} J + \sigma_{(l)}\bar{X}+ \bar{\sigma}_{(l)}{X}\in\mathfrak{sl}(2,\R)\label{Sdot}.
\end{align}
If we then insert \eref{voldot} and \eref{Sdot} back into the symplectic structure, we obtain
\begin{align}\nonumber
\Theta_{\mathcal{N}}=-\frac{1}{8\pi G}\int_{\mathcal{N}}\varepsilon\wedge\bbvar{d}\varkappa+\frac{1}{8\pi G}\int_{\mathcal{N}}\Big[&\left(\varepsilon_{ij}+\gamma^{-1}q_{ij}\right)\Omega^{-1}\mathcal{L}_l[\Omega]\,e^i\wedge\bbvar{d}(k\wedge e^j)+\\
+&\left(\varepsilon_{ij}+\gamma^{-1}q_{ij}\right)\ou{\left[P(\mathcal{L}_lS\cdot S^{-1})\right]}{i}{m}e^m\wedge\bbvar{d}(k\wedge e^j)\Big],\label{thetaN1}
\end{align}
where $P:\mathfrak{sl}(2,\R)\rightarrow\mathfrak{sl}(2,\R)$ is the projector defined in \eref{Pdef}. We now want to further simply this expression. We insert the parametrisation of $\varkappa_a$, see \eref{kappaparam}, back into \eref{thetaN1}. In addition, we also note that
\begin{align}
e^i\wedge e^j &= \varepsilon^{ij}\,\Omega^2\,d^2v_o,\\
e^i\wedge e^j\wedge\bbvar{d}k &= - e^i\wedge e^j\wedge \bbvar{d}\big(\E^{-\mu-\lambda-\rho}\di u\big)=- e^i\wedge e^j\wedge \bbvar{d}\Big[\frac{L}{2}\E^{-\mu-\lambda}\di U\Big],\label{kvar}
\end{align}
The second identity \eref{kvar} is a consequence of $\bbvar{d}m=\bbvar{f}m +\bbvar{h}\bar{m}$, see \eref{mvar}, and the parametrisation $k=-\E^{-\mu-\lambda-\rho}\di u+\zeta\bar{m}+\bar{\zeta}$ of the one-form $k_a$. The vectorial component $\zeta$ of $k_a$ is arbitrary and the value of the $\di u$ component of $k_a$ is inferred from $k_al^a=-1$. Going back to \eref{thetaN1}, we then finally obtain
\begin{align}\nonumber
\Theta_{\mathcal{N}}=&-\frac{1}{8\pi G}\int_{\partial\mathcal{N}}d^2v_o\big(\bbvar{d}\lambda -L^{-1}\bbvar{d}L\big)\Omega^2-\frac{1}{8\pi\gamma G}\int_{\mathcal{N}}d^2v_o\wedge \Omega^2\bbvar{d}\big(\varphi_{(l)}k\big)+\\\nonumber
&+\frac{1}{8\pi G}\int_{\mathcal{N}}\left[d^2v_o\wedge\bbvar{d}[\di U]\frac{\di}{\di U}\Omega^2+\frac{1}{\gamma} \ou{J}{l}{m}\ou{\left[S\cdot\bbvar{d}S^{-1}\right]}{m}{l}\,\Omega\,\di\Omega\wedge d^2v_o\right.+\\
&\hspace{6em}+\left. \Omega^2\left(\delta^m_k+\frac{1}{\gamma}\ou{J}{m}{k}\right)\ou{\left[P(\di S\cdot S^{-1})\right]}{k}{l}\ou{\left[S\cdot\bbvar{d}S^{-1}\right]}{l}{m}\wedge d^2v_o \right]\label{thetaN2}
\end{align}
To make this expression more transparent, let us introduce the abbreviations
\begin{align}
\wtilde{\Phi}& := -\varphi_{(l)}k,\label{Phidef}\\
E &:= \Omega^2,\\
\wtilde{K} & := \di U,\\
p_K & := \dot{E},\label{Pkdef}\\
\wtilde{I}&:=\frac{1}{\gamma}\Omega\,\dbar\Omega,\label{Idef}\\
\wtilde{\Pi}&:=\frac{\gamma+\I}{\gamma}\frac{L}{2}\E^{\mu-\lambda}\sigma_{(l)}\di U,\label{Pidef2}
\end{align}
where the diacritic indicates that the corresponding variable defines a density along the null rays. In addition, we have introduced in here the differential 
\begin{equation}
\dbar[\cdot] := \di U\frac{\di}{\di U}\Big[\cdot\Big].\label{dbardef}
\end{equation}
The line densities $\wtilde{I}$ and $\wtilde{\Pi}$ on the null rays $\gamma_z=\pi^{-1}(z)$ can be rearranged into the diagonal and off-diagonal components of an $SL(2,\R)$ Lie algebra-valued line density
\begin{equation}
\ou{\wtilde{\Pi}}{l}{m}  = \wtilde{I}\ou{J}{l}{m}+\wtilde{\Pi}\ou{\bar{X}}{l}{m}+\wtilde{\bar{\Pi}}\ou{X}{l}{m},\label{Pidef}\\
\end{equation}
Given these definitions, some straight-forward algebra brings the symplectic potential \eref{thetaN2} into the following form,
\begin{align}\nonumber
\Theta_{\mathcal{N}}=&-\frac{1}{8\pi G}\int_{\partial\mathcal{N}}d^2v_o\big(\bbvar{d}\lambda -L^{-1}\bbvar{d}L\big) E+\\
&+\frac{1}{8\pi G}\int_{\mathcal{N}}d^2v_o\wedge\left[p_K\bbvar{d}\wtilde{K}+\gamma^{-1}E\,\bbvar{d}\wtilde{\Phi}\,+\ou{\wtilde{\Pi}}{l}{m}\ou{\left[S\cdot\bbvar{d}S^{-1}\right]}{m}{l}\right].\label{thetaN3}
\end{align}
The basic idea to compute the resulting Poisson brackets is to introduce an auxiliary (kinematical) phase space, equipped with the symplectic potential \eref{thetaN3}, where $p_K$, $\wtilde{K}$, $E$, $\wtilde{\Phi}$, $\ou{\wtilde{\Pi}}{l}{m}$ and $S:\mathcal{N}\rightarrow SL(2,\R)$ are now treated as functionally independent coordinates on phase space. Hence, this auxiliary phase space has $2+2+2\times 3=10$ dimensions. The physical phase space is  obtained by symplectic reduction with respect to the constraints that bring us down to the two true physical degrees of freedom.

The kinematical phase space is equipped with the symplectic potential \eref{thetaN3}. It is immediate to calculate the corresponding Poisson brackets. The only non-vanishing Poisson brackets among the canonical variables $(p_K,\wtilde{K},E,\wtilde{\Phi},\ou{\wtilde{\Pi}}{l}{m},\ou{S}{l}{m})(x)$ are given by
\begin{align}
\big\{\wtilde{\Pi}(x),S(y)\big\}&=-8\pi G\, X S(y)\,\tilde{\delta}_{\mathcal{N}}(x,y),&
\big\{\wtilde{\bar{\Pi}}(x),S(y)\big\}&=-8\pi G\, \bar{X} S(y)\,\tilde{\delta}_{\mathcal{N}}(x,y),\label{Poiss1}\\
\big\{\wtilde{{\Pi}}(x),\wtilde{I}(y)\big\}&=-8\pi\I G\, \wtilde{\Pi}(y)\,\tilde{\delta}_{\mathcal{N}}(x,y),& \big\{\wtilde{\bar{\Pi}}(x),\wtilde{I}(y)\big\}&=+8\pi\I G\, \wtilde{\bar{\Pi}}(y)\,\tilde{\delta}_{\mathcal{N}}(x,y)\label{Poiss2}\\
\big\{\wtilde{\Pi}(x),\wtilde{\bar{\Pi}}(y)\big\}&=-16\pi \I G\,\wtilde{I}(y)\,\tilde{\delta}_{\mathcal{N}}(x,y),\label{Poiss3}\\
\big\{\wtilde{{I}}(x),S(y)\big\}&=+4\pi G\, J S(y)\,\tilde{\delta}_{\mathcal{N}}(x,y),\label{Poiss4}\\
\big\{p_K(x),\wtilde{K}(y)\big\} &= +8\pi G\,\tilde{\delta}_{\mathcal{N}}(x,y),\label{Poiss5}\\
\big\{E(x),\wtilde{\Phi}(y)\big\} &= +8\pi \gamma G\,\tilde{\delta}_{\mathcal{N}}(x,y),\label{Poiss6}
\end{align}
where $\tilde{\delta}_{\mathcal{N}}(x,y)$ is the Dirac distribution on $\mathcal{N}$. Notice that the first block of these equations, namely \eref{Poiss1}, \eref{Poiss2}, \eref{Poiss3}, \eref{Poiss4}, simply means that the Hamiltonian vector field of $\ou{\wtilde{\Pi}}{l}{m}$ acts as a right-invariant vector field on the group-valued configuration variable, which is $S:\mathcal{N}\rightarrow SL(2,\R)$. In fact, the commutation relations can be inferred immediately from the canonical Poisson commutation relations of $T^\ast SL(2,\R)$ equipped with its natural symplectic structure $\Theta_{{SL(2,\R)}}=\mathrm{tr}(P_S\,S\bbvar{d}S^{-1})$, with $P_S = I J + \Pi \bar{X}+\bar{\Pi}X$ denoting the $\mathfrak{sl}(2,\R)$-valued momentum in the matrix representation defined by \eref{Jgen}, \eref{Xgen} and \eref{barXgen}.

For practical reasons, it is often more useful not to work with the group-valued configuration variables, but with a Lie algebra-valued  flat $SL(2,\R)$ connection instead
\begin{equation}
\dbar{S}\cdot S^{-1}=:\wtilde{\varphi}J+\wtilde{h}\bar{X}+\wtilde{\bar{h}}X,\label{connectndef}
\end{equation}
such that
\begin{equation}
S(u,z,\bar{z}) = \mathrm{Pexp}\Big(\int_{\gamma_z(u)}\big(\wtilde{\varphi}J+\wtilde{h}\bar{X}+\wtilde{\bar{h}}X\big)\Big)S(-1,z,\bar{z}),\label{Shol}
\end{equation}
where $\mathrm{Pexp}$ is the path ordered exponential and $\gamma_z(u')$ is a null ray that starts at $u=-1$ ends at $u=u'$ and projects down onto $z\in\mathcal{C}$ on the base manifold. Inserting \eref{connectndef} back into (\ref{Poiss1}--\ref{Poiss4}), we obtain
\begin{align}
\big\{\wtilde{I}(x),\wtilde{\varphi}(y)\big\} &= 4\pi G\,\dbar_y\tilde{\delta}_{\mathcal{N}}(x,y),\\
\big\{\wtilde{I}(x),\wtilde{h}(y)\big\} &= 8\pi\I G\,\tilde{h}(y)\,\tilde{\delta}_{\mathcal{N}}(x,y),\\
\big\{\wtilde{\Pi}(x),\wtilde{\varphi}(y)\big\} &= -8\pi\I G\,\tilde{h}(y)\,\tilde{\delta}_{\mathcal{N}}(x,y),\\
\big\{\wtilde{\Pi}(x),\wtilde{h}(y)\big\} &= 0,\\
\big\{\wtilde{\Pi}(x),\wtilde{\bar{h}}(y)\big\} &= -8\pi G\,\dbar_y\tilde{\delta}_{\mathcal{N}}(x,y)-16\pi \I G\,\wtilde{\varphi}(y)\,\tilde{\delta}_{\mathcal{N}}(x,y),
\end{align}
where the notation $\dbar_y\tilde{\delta}_{\mathcal{N}}(x,y)$ indicates that the $\dbar$-differential \eref{dbardef} acts on the second argument.
\subsection{Constraints}
\noindent In this section, we introduce the constraints that reduce the kinematical phase space defined by the Poisson brackets (\ref{Poiss1}--\ref{Poiss6}) to the physical phase space of the radiative modes at the null boundary. There are five different constraints, namely the Hamiltonian, Gauss, conformal, scalar and vectorial constraints. We introduce them below.\medskip

\emph{(i) Gauss constraint.} First of all, we have the Gauss constraint. Let $\Lambda:\mathcal{N}\rightarrow\R$ be a test function of compact support. To impose \eref{Idef} on the kinematical phase space (\ref{Poiss1}--\ref{Poiss6}), we introduce the constraint
\begin{equation}
\forall\Lambda : G[\Lambda] = \int_{\mathcal{N}}d^2v_o\wedge\Lambda\left(\wtilde{I}-\frac{1}{2\gamma}\dbar E\right)\stackrel{!}{=}0,\label{Gausscons}
\end{equation}
we will see below that this constraint is first-class and generates local $U(1)$ gauge transformations along the null boundary.\medskip

\emph{(ii) Conformal constraint.} Next, we have a conformal constraint that imposes \eref{Pkdef} via
\begin{equation}
\forall \lambda:C[\lambda] = \int_{\mathcal{N}}d^2v_o\wedge\lambda\left(p_K\wtilde{K}-\dbar{E}\right)\stackrel{!}{=}0,\label{Ccons}
\end{equation}
where $\lambda:\mathcal{N}\rightarrow\R$ is a test function of compact support. Notice that $p_K\wtilde{K}$ is a squeeze operator.\medskip

\emph{(iii) Scalar constraint.} At the kinematical level, the $SL(2,\R)$ holonomy \eref{Shol} and the $\mathfrak{sl}(2,\R)$-valued momentum \eref{Pidef} are functionally independent. The functional dependence is obtained by imposing \eref{Sdot} as a constrained on phase space. This involves a scalar constraint and a complex-valued (momentum) constraint. The scalar constraint is given by
\begin{equation}
\forall \mu: D[\mu]=\int_{\mathcal{N}}d^2v_o\wedge \mu\left(\tilde{\Phi}-\tilde{\varphi}\right)\stackrel{!}{=}0,\label{Dcons}
\end{equation}
where $\mu:\mathcal{N}\rightarrow\R$ is a test function and $\tilde{\Phi}$ and $\tilde{\phi}$ are defined in \eref{connectndef} and \eref{Phidef} respectively.\medskip

\emph{(iv) Vectorial constraint.} Besides the scalar constraint, we also need to impose the vectorial component of \eref{Sdot}. Equation \eref{connectndef} is compatible with \eref{Sdot} only if we impose
\begin{equation}
\forall\zeta:V[\bar{\zeta}]=\int_{\mathcal{N}}d^2v_o\wedge\bar{\zeta}\E^{-2\I\Delta}\left(\Omega^{-1}\wtilde{\Pi}-\frac{\gamma+\I}{\gamma}\Omega\wtilde{h}\right) \stackrel{!}{=}0,\label{Vcons}
\end{equation}
where $\zeta:\mathcal{N}\rightarrow\C$ is a test function of compact support and $\Delta$ is the $U(1)$ angle
\begin{equation}
\Delta(u,z,\bar{z}) = \int_{\gamma_z(u)}\wtilde{\varphi},\label{U1hol}
\end{equation}
with $\gamma_z\subset\mathcal{N}$ denoting a null ray that starts at $u=-1$, ends at $u=u'$ and projects down onto $z\in\mathcal{C}$ on the base manifold. Strictly speaking, this $U(1)$ dressing of the constraint is not necessary, but we will see below that it greatly simplifies the constraint analysis. Notice also that $V[\bar{\zeta}]$ is a complex constraint, hence we have two real constraints $\operatorname{\mathfrak{Re}}(V[\bar{\zeta}])=0$ and $\operatorname{\mathfrak{Im}}(V[\bar{\zeta}])=0$ to impose. All other constraints are real. Notice also that the vector constraints \eref{Vcons} are essentially creation and annihilation operators with the conformal factor playing the role of the characteristic length scale of the oscillators. \medskip

\emph{(v) Hamiltonian constraint.} The Raychaudhuri equation determines the evolution of the conformal factor along the null generators. Going back to \eref{Cxi1} and \eref{Cxi4}, we obtain
\begin{equation}
\mathcal{L}_l\left[\vartheta_{(l)}\right]-\kappa_{(l)}\vartheta_{(l)}=-\frac{1}{2}\vartheta_{(l)}^2-2\sigma_{(l)}\bar{\sigma}_{(l)}.
\end{equation}
The expression simplifies once we evaluate it for the vector field $U^a$, defined in \eref{Uvecdef}. Taking into account that its expansion and non-affinity are related by \eref{Unonaff}, the Raychaudhuri equation turns into the constraint
\begin{equation}
\frac{1}{2}\frac{\di^2}{\di U^2}\Omega^2 + \sigma\bar{\sigma}=0,\label{Raycons}
\end{equation}
where
\begin{equation}
\sigma(u,z,\bar{z}) = \frac{L(z,\bar{z})}{2}\E^{-\lambda(u,z,\bar{z})}\sigma_{(l)}(u,z,\bar{z}),
\end{equation}
is the rescaled shear. To realise this equation as a constraint on the kinematical phase space (\ref{Poiss1}--\ref{Poiss6}), consider first the following vertical vector field
\begin{equation}
\xi^a_N = N\partial^a_u,
\end{equation}
where $\partial^a_u$ is the fiducial vector field introduced in \eref{lparam} and $N:\mathcal{N}\rightarrow \R$ is a lapse function that satisfies the boundary conditions
\begin{equation}
N\big|_{\mathcal{C}_0}=N\big|_{\mathcal{C}_1}=0.
\end{equation}
Consider then the Lie derivative with respect to this vector field such that we are able to define the generator 
\begin{equation}
H[N]:=\Theta_{\mathcal{N}}(\mathcal{L}_{\xi_N}).\label{Hdef1}
\end{equation}
It is straightforward to check that for every $N$ of compact support this generator vanishes provided the constraints (\ref{Gausscons}--\ref{Vcons}), \eref{Raycons} are satisfied. Indeed,

\begin{align}
\nonumber H[N] & = \frac{1}{8\pi G}\int_{\mathcal{N}}d^2v_o\wedge \Big[-\frac{1}{\gamma}\mathcal{L}_{\xi_N}[E]\wtilde{\Phi}-\mathcal{L}_{\xi_N}[p_K]\wtilde{K}+\ou{\wtilde{\Pi}}{l}{m}\ou{\left[S\cdot\mathcal{L}_{\xi_N}S^{-1}\right]}{m}{l}\Big]\approx\\
\nonumber &\approx \frac{1}{8\pi G}\int_{\mathcal{N}}d^2v_o\wedge N\bigg[-\frac{1}{\gamma}\wtilde{\Phi}\frac{\di}{\di u}\Omega^2-\di U\frac{\di}{\di u}\frac{\di}{\di U}\Omega^2+2\,\wtilde{I}\partial_u\hook\wtilde{\varphi}
-\left(\wtilde{\Pi}\,\partial_u\hook\wtilde{\bar{h}}+\CC\right)\bigg]\approx\\
\nonumber&\approx\frac{1}{8\pi G}\int_{\mathcal{N}}d^2v_o\wedge N\bigg[-\di U (\partial_uU)\frac{\di^2}{\di U^2}\Omega^2-\left(\frac{\gamma}{\gamma-\I}\Omega^{-2}\wtilde{\Pi}\,\partial_u\hook\wtilde{\bar{\Pi}}+\CC\right)\bigg]\approx\\
\nonumber&\approx\frac{1}{8\pi G}\int_{\mathcal{N}}d^2v_o\wedge\di U\,N(\partial_uU)\bigg[-\frac{\di^2}{\di U^2}\Omega^2-\frac{L^2}{4}\E^{-2\lambda}\left(\frac{\gamma+\I}{\gamma}\sigma_{(l)}\bar{\sigma}_{(l)}+\CC\right)\bigg]\approx\\
&\approx-\frac{1}{4\pi G}\int_{\mathcal{N}}d^2v_o\wedge\di U\,\mathcal{L}_{\xi_N}[U]\bigg[\frac{1}{2}\frac{\di^2}{\di U^2}\Omega^2+\sigma\bar{\sigma}\bigg]\approx 0,\label{Hdef2}
\end{align}
where $\approx$ stands for equality on the constraint hypersurface and $\partial_u\hook$ denotes the interior product, e.g.\ $\partial_u\hook \di U =\partial_uU$.

\subsection{Dirac analysis of the constraint algebra}
\noindent It is immediate to check that the Hamiltonian constraint $H[N]$ is the generators of vertical diffeomorphisms $\varphi_N=\exp{\xi_N}$ on the kinematical phase space defined by the commutation relations (\ref{Poiss1}--\ref{Poiss6}). Going back to the definition of the constraints, we find, in fact,
\begin{align}
\big\{H[N],H[N']\big\}&=-H\left[N\frac{\di}{\di u}N'-N'\frac{\di}{\di u}N\right],\label{alge1}\\
\big\{H[N],G[\Lambda]\big\} &= -G\left[N\frac{\di}{\di u}\Lambda\right],\label{alge2}\\
\big\{H[N],C[\lambda]\big\} &= -C\left[N\frac{\di}{\di u}\lambda\right],\label{alge3}\\
\big\{H[N],D[\mu]\big\} &= -D\left[N\frac{\di}{\di u}\mu\right],\label{alge4}\\
\big\{H[N],V[\bar{\zeta}]\big\} &= -V\left[N\frac{\di}{\di u}\zeta\right].\label{alge5}
\end{align}

The Poisson brackets \eref{Poiss2}, \eref{Poiss4} imply that $\tilde{I}$ is the generator of an $U(1)$ subgroup of $SL(2,\R)$. Going back to the definition of the Guess constraint \eref{Gausscons}, it is then immediate to infer the commutation relations
\begin{align}
\big\{G[\Lambda],G[\Lambda']\big\}&=0,\label{alg6}\\
\big\{G[\Lambda],C[\lambda]\big\}&=0,\label{alg7}\\
\big\{G[\Lambda],D[\mu]\big\}&=0,\label{alg8}\\
\big\{G[\Lambda],V[\bar{\zeta}]\big\}&=0.\label{alg9}
\end{align}
In proving (\ref{alg6}--\ref{alg9}) the only  difficulty arises with equation \eref{alg9}, where we have to take into account that the constraint has been dressed by the introduction of an $U(1)$ holonomy, see equation \eref{Vcons} and \eref{U1hol}.

All other constraints are second class. A few entries of the corresponding Dirac matrix vanish identically,
\begin{align}
\big\{C[\lambda],C[\lambda']\big\} & = 0,\label{alg10}\\
\big\{C[\lambda],V[\bar{\zeta}]\big\} & = 0,\label{alg11}\\
\big\{D[\mu],D[\mu']\big\} & = 0.\label{alg12}
\end{align}
Since $\wtilde{\Phi}$ and $E$ are conjugate, but $p_K\wtilde{K}$ and $\wtilde{\varphi}$ Poisson commutes with both $\wtilde{\Phi}$ and $E$, we obtain that $C$ and $D$ are conjugate,
\begin{equation}
\big\{C[\lambda],D[\mu]\big\}=-8\pi\gamma G\int_{\mathcal{N}}d^2v_o\wedge\lambda\dbar\mu.\label{alg13a}
\end{equation}
We then also find
\begin{equation}
\big\{D[\mu],V[\bar{\zeta}]\big\}=4\pi G\int_{\mathcal{N}}d^2v_o\wedge\mu\,\bar{\zeta}\,\E^{-2\I\Delta}\,\Omega^{-1}\left[(\gamma-\I)\wtilde{h}+\Omega^{-2}\wtilde{\Pi}\right].\label{alg13b}
\end{equation}
If the vector constraint \eref{Vcons} is satisfied, the right hand side turns into
\begin{equation}
\big\{D[\mu],V[\bar{\zeta}]\big\}\approx8\pi\gamma G\int_{\mathcal{N}}d^2v_o\wedge \mu\,\bar{\zeta}\E^{-2\I\Delta}\Omega^{-1}\wtilde{h},
\end{equation}
where, once again, $\approx$ stands for \emph{equal on the constraint hypersurface}. If the shear vanishes, hence $\tilde{h}\approx \Omega^{-1}\sigma\di U=0$, the constraints $D[\mu]=0$ and $V[\bar{\zeta}]=0$ commute under the Poisson bracket.

We are now left to compute the Poisson brackets among $V$ and $\bar{V}$. First of all, we have
\begin{align}
\big\{V[\bar{\zeta}],V[\bar{\zeta}']\big\}=\int_{\mathcal{N}}d^2v_o\wedge\bigg[\bar{\zeta}'\big\{V[\bar{\zeta}],\E^{-2\I\Delta}\big\}\Big(\Omega^{-1}\wtilde{\Pi}-\frac{\gamma+\I}{\gamma}\Omega\wtilde{h}\Big)-(\zeta\leftrightarrow\zeta')\bigg].
\end{align}
Therefore,
\begin{equation}
\big\{V[\bar{\zeta}],V[\bar\zeta']\big\}\approx 0.\label{alg14}
\end{equation}
Finally, we also need to evaluate $\{V,\bar{V}\}$. A straight forward calculation yields
\begin{align}
\nonumber\{V[\bar{\zeta}],\bar{V}[{\zeta}']\big\}  = &\int_{\mathcal{N}}d^2v_o\wedge\bigg[\zeta'\big\{V[\bar\zeta],\E^{+2\I\Delta}\big\}\Big(\Omega^{-1}\wtilde{\bar{\Pi}}-\frac{\gamma-\I}{\gamma}\Omega\wtilde{\bar h}\Big)+\\
\nonumber &\hspace{6.1em}-\bar{\zeta}\big\{\bar{V}[\zeta'],\E^{-2\I\Delta}\big\}\Big(\Omega^{-1}\wtilde{\Pi}-\frac{\gamma+\I}{\gamma}\Omega\wtilde{h}\Big)\bigg]+\\
\nonumber&+16\pi G\int_{\mathcal{N}}d^2v_o\wedge\bigg[-\I\bar\zeta\zeta'\Omega^{-2}\wtilde{I}+\frac{\I}{\gamma}\bar{\zeta}\zeta'\Omega^{-1}\dbar\Omega-\bar{\zeta}\dbar\zeta'\bigg].
\end{align}
Going back to the definition of the Gauss constraint \eref{Gausscons} and the definition of the the vectorial constraint \eref{Vcons}, we thus see
\begin{equation}
\{V[\bar{\zeta}],\bar{V}[{\zeta}']\big\} \approx-16\pi G\int_{\mathcal{N}}d^2v_o\wedge\bar{\zeta}\dbar\zeta',\label{alg15}
\end{equation}
where $\approx$ denotes equality on the constraint hypersurface.\medskip

Let us briefly summarise. The kinematical phase space is defined by the Poisson commutation relations (\ref{Poiss1}--\ref{Poiss6}). The physical phase space is obtained by imposing the constraints. There are both first-class and second-class constraints. The Hamiltonian (Raychaudhuri equation) and Gauss constraint  are first class, all other constraints (\ref{Ccons}, \ref{Dcons}, \ref{Vcons}) are second-class. The kinematical phase space, which is equipped with the Poisson brackets (\ref{Poiss1}--\ref{Poiss6}), is ten-dimensional. The resulting physical phase space is two-dimensional and spanned by the Dirac observables of the theory. In the next section, we compute the Poisson brackets between such observables.
\subsection{Dirac observables from $SL(2,\R)$ holonomies}
\noindent Finally, let us explain how to identify coordinates (Dirac observables) on the physical phase space. The physical phase space is equipped with the Dirac bracket. Let $\{\Phi_\alpha\}_\alpha$ be a family of second-class constraints and $\{\Phi_\alpha,\Phi_\beta\}=\Delta_{\alpha\beta}: \det\Delta\neq 0$ be the Dirac matrix  and $\delta^{\alpha\beta}:\Delta^{\alpha\mu}\Delta_{\mu\beta}=\delta^\alpha_\beta$ be its inverse. The resulting Dirac bracket is defined by removing the auxiliary directions from phase space, i.e.\
\begin{equation}
\big\{A,B\big\}^\ast = \big\{A,B\big\}- \big\{A,\Phi_\alpha\big\}\Delta^{\alpha\beta}\big\{\Phi_\beta,B\big\},\label{dirac0}
\end{equation}
where $\alpha$ is some unspecified multi-index. Summation (integration) is understood over all repeated (De\,Witt) indices $\alpha,\beta,\dots$. 

We now want to compute the Dirac bracket for a specific class of partial observables, namely the matrix entries of the $SL(2,\R)$ holonomy \eref{Shol}. The calculation is straight forward: first of all, we note that the matrix entries of the $SL(2,\R)$ holonomy \eref{Shol} commute among themselves with respect to the Poisson brackets of the kinematical phase space. In other words,
\begin{equation}
\big\{\ou{S}{i}{m}(u,z,\bar{z}),\ou{S}{j}{n}(u',z',\bar{z}')\big\}=0.
\end{equation}
Furthermore,
\begin{align}
\big\{C[\lambda],\ou{S}{i}{m}(u,z,\bar{z})\big\}&=0,\\
\big\{D[\mu],\ou{S}{i}{m}(u,z,\bar{z})\big\}&=0,
\end{align}
for all test functions $\lambda:\mathcal{N}\rightarrow \R$ and $\mu:\mathcal{N}\rightarrow \R$. The only entries of the inverse Dirac matrix that we need to evaluate in order to calculate $\{S,S'\}^\ast$ are therefore those that correspond to the vector constraint. Schematically,
\begin{equation}
\big\{S,S'\big\}^\ast = -\big\{S,V_\alpha\}\Delta^{\alpha\bar{\alpha}}\big\{\bar{V}_{\bar\alpha},S'\}+\CC
\end{equation}
Inverting the right hand side of \eref{alg15}, we obtain
\begin{align}\nonumber
\big\{\ou{S}{i}{m}(u,z,\bar{z}),&\,\ou{S}{j}{n}(u',z',\bar{z}')\big\}^\ast = \frac{1}{16\pi G}\int_{-1}^1\di u_1\int_{-1}^1 \di u_o\int_{\mathcal{C}}[d^2v_o]_{(w,\bar{w})}\bigg[\\
&\times\big\{\ou{S}{i}{m}(u,z,\bar{z}),V(u_1,w,\bar{w})\big\}\Theta(u_1-u_o)\big\{\bar{V}(u_o,w,\bar{w}),\ou{S}{j}{n}(u',z',\bar{z}'),\big\}+\CC\bigg],\label{dirac1}
\end{align}
where $\Theta(u)$ is the distribution
\begin{equation}
\Theta(u)=\begin{cases}+\frac{1}{2}, u>0,\\ -\frac{1}{2}, u\leq 0,\end{cases}
\end{equation}
and $V(u,z,\bar{z})$ is the integrand that appears in the definition \eref{Vcons} of the vector constraint, i.e.\
\begin{equation}
V[\bar{\zeta}]\equiv\int_{-1}^1\di u\int_{\mathcal{C}}[d^2v_o]_{(z,\bar{z})}\,\bar{\zeta}(u,z,\bar{z})\,{{V}}(u,z,\bar{z}) \stackrel{!}{=}0.
\end{equation}

Equation \eref{dirac1} provides an important intermediate step, because it tells us that we know the Dirac bracket $\{S,S'\}^\ast$, if we know the ordinary Poisson bracket $\{V,S\}$
on the kinematical phase space. Given the fundamental Poisson brackets (\ref{Poiss1}--\ref{Poiss6}) and the definition of the vector constraint \eref{Vcons}, the Poisson bracket $\{V,S\}$ is immediate to calculate. We obtain
\begin{equation}
\big\{V[\bar{\zeta}],\ou{S}{i}{m}(x)\big\} = -8\pi G\, \bar{\zeta}(x)\,\E^{-2\I\Delta(x)}\,\Omega^{-1}(x)\,\ou{X}{i}{j}\ou{S}{j}{m}(x),\label{dirac2}
\end{equation}
where $x\equiv(u,z,\bar{z})$ labels a point on $\mathcal{N}$. Inserting \eref{dirac2} back into the expression for the Dirac bracket \eref{dirac1}, we infer the commutation relations for the $SL(2,\R)$ holonomy
\begin{align}\nonumber
\big\{\ou{S}{i}{m}(x),\ou{S}{j}{n}(y)\big\}^\ast = -4\pi G\,\Theta(x,y)&\,\delta^{(2)}(x,y)\,\Omega^{-1}(x)\,\Omega^{-1}(y)\\
&\times\bigg[\E^{-2\,\I\,(\Delta(x)-\Delta(y))}\ou{\big[XS(x)\big]}{i}{m}\ou{\big[\bar{X}S(y)\big]}{j}{n}+\CC\bigg].\label{dirac3}
\end{align}
Let us briefly summarise our notation in this equation: $\ou{S}{i}{j}(x)$ is the $SL(2,\R)$ holonomy that determines the conformal class of the co-dyad $\ou{e}{i}{a}$ on the null boundary, $\Theta(x,y)$ is the step function $\Theta(x,y)=\Theta(u(x)-u(y))=\pm\frac{1}{2}$, and $\delta^{(2)}(x,y)\equiv \delta_{\mathcal{C}}(\pi(x),\pi(y))$ is the two-dimensional Dirac distribution on the base manifold. In addition, $\Omega$ is the conformal factor \eref{conffactr}, $\E^{\I\Delta}$, defined via equations \eref{U1hol} and \eref{connectndef},  is the $U(1)$ holonomy along the null generators, and $\ou{X}{i}{j}=m^im_j$ is the complex-valued $SL(2,\R)$ translation generator that we introduced in \eref{Xgen} above.

The physical phase space has two local degrees of freedom. The $SL(2,\R)$ holonomy has three functionally independent components and it is not quite a Dirac observable yet. In fact, it does not commute with the first-class constraints. Instead, we have
\begin{align}
\big\{H[N],\ou{S}{i}{k}(u,z,\bar{z})\big\}&= N(u,z,\bar{z})\frac{\di}{\di u}\ou{S}{i}{k}(u,z,\bar{z}),\\
\big\{G[\Lambda],\ou{S}{i}{k}(u,z,\bar{z})\big\} & = 4\pi G\,\Lambda(u,z,\bar{z})\, \ou{J}{i}{m}\ou{S}{m}{k}(u,\bar{z},\bar{z}).
\end{align}
Imposing the Gauss constraint amounts to computing the quotient $SL(2,\R)/U(1)$. An example for such an $U(1)$-invariant observable is  given by the $U(1)$ dressed $SL(2,\R)$ holonomy,  
\begin{equation}
H =\E^{-J\Delta }S,\label{dressedS}
\end{equation}
where $J$ denotes the complex structure \eref{Jgen} and $\E^{\I\Delta}$ is the $U(1)$ holonomy defined via \eref{U1hol}. Another natural example is the two-dimensional conformally rescaled  metric,
\begin{equation}
\gamma_{ij}:= q_{mn}\ou{S}{m}{i}\ou{S}{n}{j},
\end{equation}
where $q_{mn}=\delta_{mn}$ is the two-dimensional internal metric \eref{intrnlq} on the auxiliary internal vector space $\mathbb{V}$. Notice that both $H$ and $\gamma$ are still only partial observables, because neither of them commute with the Hamilton constraint. True Dirac observables can be  introduced following standard constructions. An example is given by
\begin{equation}
\gamma_{ij}(T,z,\bar{z})=\int_{-1}^{+1}\di u\, (\partial_u\Omega)(u,z,\bar{z})\, \delta\big(T-\Omega(u,z,\bar{z})\big)\,\gamma_{ij}(u,z,\bar{z}),\label{Dobs}
\end{equation}
which is the functional on phase space that evaluates the conformally rescaled metric $\gamma_{ij}$ at those points of the null ray $\gamma_z=\pi^{-1}(z)$, where the conformal factor takes the value $T$. Hence the area serves as a relational clock---Bondi's luminosity distance---to define the Dirac observable \eref{Dobs}. Since the conformal factor $\Omega$ commutes with all second-class constraints except $D[\mu]$, and since the  inverse Dirac matrix has neither diagonal {$D\times D$} entries nor any off-diagonal {$D\times C$} ($C\times D$) entries, the commutation relations for $\gamma_{ij}$ can be immediately inferred from \eref{dirac1}. A short calculation yields
\begin{equation}
\big\{\gamma_{ij}(T,z,\bar{z}),\gamma_{kl}(T',z',\bar{z}')\big\}^\ast = -16\pi G\, \Theta(T-T')\,\delta^{(2)}(z,z')\Big(\Xi_{ij}(T,z,\bar{z})\,\Xi_{kl}(T',z,\bar{z})+\CC\Big),
\end{equation}
where $\Xi_{ij}$ is an abbreviation for
\begin{equation}
\Xi_{ij} = \Omega^{-1}m_km_l\ou{H}{k}{i}\ou{H}{l}{j}.
\end{equation}

Notice that the Dirac observable \eref{Dobs} is constructed using a \emph{line dressing}, where a kinematical (gauge variant, partial) observable is integrated along a light ray to obtain a gauge invariant observable. A recent discussion of such dressed observables in the context of perturbative gravity can be found in \cite{Giddings:2019hjc,Donnelly:2018nbv}.

\section{Summary and Outlook}
\noindent \emph{Summary.} Let us briefly summarise. 
The first set of results were developed in \hyperref[sec2]{Section 2}, where we considered the Palatini action with the parity odd Holst term $F_{\alpha\beta}[A]\wedge e^\alpha\wedge e^\beta$ in a finite domain with a boundary that is null. We specified the  boundary conditions and introduced the corresponding boundary term along the null surface. The boundary conditions \eref{gravitondef} are such that a gauge equivalence class of boundary fields is kept fixed. This equivalence class is characterised by two local degrees of freedom along the null surface, and additional corner data---gravitational edge modes \cite{Balachandran:1994up,Strominger:1997eq,Banados:1998ta,Carlip:2005zn,Afshar:2016wfy,Compere:2017knf,Wieland:2018ymr,Namburi:2019qja,Wieland_2020,Freidel:2020xyx,Rovelli:2020mpk} at the two boundaries of $\mathcal{N}$. We introduced the pre-symplectic potential on the quasi-local radiative phase space, and we identified the gauge symmetries of the resulting pre-symplectic two-form. An important observation was made in equation \eref{Kzero} and \eref{U1gen}, where we saw that the Barbero\,--\,Immirzi parameter deforms the $U(1)$ frame rotations (\ref{U1kappa}, \ref{U1l}, \ref{U1m}). The resulting $U(1)$ charge \eref{Ldef} is proportional to the generator of  dilations \eref{Kdef}, which is itself a multiple of the cross-sectional area. This deformation is responsible for the quantisation of area in loop quantum gravity, where the Barbero\,--\,Immirzi parameter $\gamma$  determines the relation \eref{KLcons} between dilatations and $U(1)$ frame rotations on the null cone. 

In the second part of the paper, \hyperref[sec2]{Section 2}, we studied the Dirac observables on the null surface. To infer the Poisson brackets, we found it useful to embed the covariant phase space into a kinematical phase space that contains an auxiliary $SL(2,\R)$ holonomy flux algebra (\ref{Poiss1}-\ref{Poiss3}). The physical phase space is obtained by symplectic reduction with respect to first-class and second-class constraints. The two first-class constraints are the generators  of vertical diffeomorphisms and $U(1)$ frame rotations, namely the Hamilton constraint \eref{Hdef2} and the Gauss constraint \eref{Gausscons}. The second class constraints impose the conformal matching \eref{Ccons} and additional constraints on the $\mathfrak{sl}(2,\R)$ currents, i.e.\ equations \eref{Dcons} and \eref{Vcons}. We computed the constraint algebra and defined the Dirac bracket \eref{dirac0}. An important result was given in equation \eref{dirac3}, were we found the commutation relations among the $SL(2,\R)$ holonomy. At null infinity, its matrix elements describe both gravitational radiation as well as gravitational memory \cite{PhysRevLett.67.1486,memory_frauendiener}. Finally, we introduced a line dressing \cite{Giddings:2019hjc,Donnelly:2018nbv} to construct physical observables, which are invariant under  $U(1)$ gauge transformations and vertical diffeomorphisms.

\medskip

\noindent \emph{Outlook.}  It seems an extremely daunting task to quantise the Dirac bracket \eref{dirac3} directly. The right hand side \eref{dirac3} involves not only the $SL(2,\R)$ holonomies $\ou{S}{i}{j}$, but also the conformal factor. Yet the conformal  factor depends via the Raychaudhuri equation \eref{Raycons} on the shear $\sigma$, which is itself obtained from the $SL(2,\R)$ Maurer--Cartan form $S^{-1}\dbar S$, see \eref{Sdot} and \eref{connectndef}. The entire construction is highly non-linear. Various truncations seems necessary to quantise the algebra  in an effective way. Let us briefly consider three possible such truncations below.
\begin{description}
\item[- Asymptotic quantisation:] The first and perhaps simplest possibility is to restrict ourselves to asymptotically flat spacetimes ($\Lambda=0$) and quantise the asymptotic phase space alone. Consider an auxiliary double null foliation $\{\mathcal{N}_{r}\}_{r> 0}$ in a vicinity of future (past) null infinity $\mathcal{I}^\pm$ as in \cite{Wieland:2020gno} such that $\mathcal{I}^\pm=\lim_{r\rightarrow\infty}\mathcal{N}_{r}$. Given such a foliation, the radiative phase space can be obtained  from an asymptotic $r\rightarrow\infty$ limit of the quasi-local radiative phase space introduced above. 
If we take the falloff conditions for the various spin coefficients of an adapted Newman--Penrose tetrad into account, we obtain the $1/r$ expansion
\begin{align}
\Omega(u,r,z,\bar{z}) & = r\,\Omega^{(0)}(z,\bar{z})+ \mathcal{O}(r^0),\label{falloff1}\\
\ou{H}{i}{j}(u,r,z,\bar{z}) & = \left[\delta^i_k +\frac{\sigma^{(0)}(u,z,\bar{z})}{\Omega^{(0)}(z,\bar{z})}\frac{1}{r}\ou{\bar{X}}{i}{k}+\frac{\bar{\sigma}^{(0)}(u,z,\bar{z})}{\Omega^{(0)}(z,\bar{z})}\frac{1}{r}\ou{{X}}{i}{k}+\mathcal{O}(r^{-2})\right] \ou{[S_o]}{k}{j}(z,\bar{z}),\label{falloff2}
\end{align}
where $\ou{H}{i}{j}$ is the dressed $SL(2,\R)$ holonomy \eref{dressedS}, $\Omega$ is the (inverse) conformal factor and $S_o(z,\bar{z})\in SL(2,\R)$ is the initial condition that appears in  \eref{Shol}. In the asymptotic $\Omega\rightarrow\infty$ limit, only the first two terms of the $1/r$ expansion of the dressed $SL(2,\R)$ holonomy enter the symplectic potential \eref{thetaN3}. For  $r\rightarrow\infty$, the $SL(2,\R)$ group-valued commutation relations \eref{dirac3} turn into the Lie algebra-valued commutation relations for the asymptotic shear. The result of this contraction on phase space defines the Poisson brackets
\begin{equation}
\big\{\sigma^{(0)}(x),\bar{\sigma}^{(0)}(y)\big\}^\ast = -4\pi G\,\Theta(x,y)\,\delta^{(2)}(x,y)\label{asymptcomm}
\end{equation}
of the radiative data at null infinity.
Seperating \eref{asymptcomm} into positive and negative frequency modes, we obtain an infinite set of harmonic oscillators. The Fock vacuum is then simply the state that is annihilated by the positive frequency part of the asymptotic shear $\sigma_{ab}^{(0)}=\sigma^{(0)}\bar{m}_a\bar{m}_b+\CC$ 

From the perspective of the quasi-local phase space, we find such an asymptotic quantisation problematic. The conformal factor is treated as a classical field and sent to infinity before considering the quantum theory.\footnote{In addition, the $SL(2,\R)$ zero-mode $S_o(z,\bar{z})$ is usually ignored as well. An approach, where such $SL(2,\R)$ edge modes are not ignored, can be found in \cite{Freidel:2020svx,DePaoli:2017sar}.} This seems to be in conflict with the Heisenberg uncertainty relations that tell us that the area two-form is canonically conjugate to a $U(1)$ gauge parameter, see \eref{U1gen}. Hence we can not impose simultaneous gauge conditions on both the $U(1)$ gauge parameter $\tilde{\varphi}=\mathcal{O}(r^{-1})$ and the conformal factor $\Omega=\mathcal{O}(r)$. A possible solution is to quantise the quasi-local phase space at the full non-perturbative level and construct coherent states that are peaked at appropriate values of $\Omega$ and $\tilde{\varphi}$. Once such coherent states are found, we can define asymptotic states by sending $\Omega\rightarrow\infty$. Even if such a construction succeeds, the task remains how to recover the ordinary $S$ matrix in this framework. In fact, the $S$ matrix is a highly entangled boundary state (linear functional) \cite{Oeckl2003,Oeckl:2016tlj} on the tensor product Hilbert space $\mathcal{H}_+\otimes\mathcal{H}_-^\dagger$, where $\mathcal{H}_\pm$ is the Fock space for future (past) null infinity.  Defining this tensor product amounts to join the two asymptotic boundaries into a single Hilbert space. A rigorous construction of such a joint Hilbert space may very well be possible only via an entangling product \cite{Donnelly:2016auv,Wong:2017pdm} that takes into account that the gravitational edge modes, which are located at a joint cross section of the two asymptotic boundaries (such as $i_o$), must be quantised as well.
\item[- Loop quantization:] In loop gravity \cite{zakolec,status}, we face the opposite problem. We have a straight-forward and rigorous quantisation of gravitational edge modes, see e.g.\ \cite{Wieland:2017cmf,Freidel:2020svx}, but it is difficult to characterise the two radiative modes. Consider, for example a spin network state on a partial Cauchy hypersurface. Such a state consists of superpositions of gravitational Wilson lines. Every such Wilson line represents a quantum of area. At the boundary of the partial Cauchy surface, the Wilson lines terminate and a quantum of area is excited. If the resulting boundary state on a cross section $\mathcal{C}$ of the null surface is an eigenstate of the area operator, we would have, e.g.\
\begin{equation}
\widehat{\Omega}^2(z,\bar{z})|\vec{\jmath},\vec{m},\mathcal{C}\rangle = 8\pi\,\hbar\gamma G/c^3\,\sum_k \sqrt{j_k(j_k+1))}\,\delta^{(2)}(\bar{z}_k,z_k;z,\bar{z})|\vec{\jmath},\vec{m},\mathcal{C}\rangle,\label{horizonstate}
\end{equation}
where $\vec{\jmath},\vec{m}$ are the usual $SU(2)$ spin labels. 
Hence the conformal factor is quantised. The question is then to construct a representation of the shear at the quantum level, such that we can use the Hamiltonian constraint \eref{Raycons} to generate the evolution of the boundary states along the null generators.  Twisted geometries, which admit a representations of the relevant $SL(2,\R)$ degrees of freedom \cite{twistedconn}, may provide a natural framework to study this problem in the spin network representation. In three-dimensional euclidean gravity, the problem of how to explain such boundary evolution for spin network states has been studied in \cite{Dittrich:2018xuk,Dittrich:2017hnl}.

\item[- Other truncations:] It would be highly desirable to develop yet another truncation, where we only consider a finite-dimensional symplectic subspace of the quasi-local phase space that we defined in above. Such a subspace could be found by various UV cutoffs. If, for example, the cross sections are spherical, we could choose an auxiliary (round) metric, decompose all configuration variables into spherical harmonics (with respect to the auxiliary metric), and introduce an effective UV cutoff for large spins and frequencies, see e.g.\ \cite{Wieland:2017cmf}. 
It would be very useful to quantise such a finite-dimensional phase-space, and define different  UV completions via different refining maps \cite{Sahlmann:2001nv,Dittrich:2014ala} that embed the finite-dimensional Hilbert spaces into larger and larger tensor product spaces. 
\end{description}

\paragraph{- Acknowledgments} Support from the Institute for Quantum Optics and Quantum Information is gratefully acknowledged. 
This research was supported in part by the ID 61466 grant from the John Templeton Foundation, as part of The Quantum Information Structure of Spacetime (QISS) Project (qiss.fr). The opinions expressed in this publication are those of the author and do not necessarily reflect the views of the John Templeton Foundation.


\providecommand{\href}[2]{#2}\begingroup\raggedright\endgroup


\end{document}